\documentclass[a4paper]{aipproc}
\usepackage{graphicx,subfig}

\layoutstyle{6x9}

\SetInternalRegister\hbadness{8000} 

\newcommand\doingARLO[2][]{%
  \ifx\mmref\undefined #1\else #2\fi
}

\begin{document}
\title 
      [Constraining dark energy]
      {Geometrical constraints on dark energy models}


\author{Ruth Lazkoz}{
  address={Fisika Teorikoa, Euskal Herriko Unibertsitatea, 644 posta kutxatila, 48007 Bilbao, Espa\~na},
  email={ruth.lazkoz@ehu.es},
}
\copyrightyear  {2007}

\begin{abstract}
This contribution intends to give a pedagogical introduction to the topic of dark energy (the mysterious agent supposed to drive the observed late time acceleration of the Universe) and to various observational tests which require only assumptions on the geometry of the Universe. Those tests are the supernovae luminosity, the CMB shift, the direct Hubble data, and the
baryon acoustic oscillations test. An historical overview of Cosmology is followed by some generalities on FRW spacetimes (the best large-scale description of the Universe), and then the test themselves are discussed. A convenient section on statistical inference is included as well.
\end{abstract}

\date{\today}

\maketitle
\section{Introduction}
Cosmology is a branch of physics which is experiencing a tremendously fast development lately triggered by the arrival 
of many new observational data of ever more exquisite precision. These findings  have been crucial for the improvement in  our understanding of the Universe, but the vast amount of knowledge on our Cosmos which we have nowadays would never have been possible without the concerted effort by experimentalists and theoreticians.
One of the results of this fantastic intellectual pursue is the puzzling discovery that there is in the Universe a manifestation of the repulsive side of gravity. This is the topic to which this lectures are devoted, and more specifically I wish to address some methods which the community believes are useful for building a deeper understanding of why, how and when our universe began to accelerate. Put in  more modest words,  these lectures will dissert about how one can take advantage of various observational datasets to describe some basic features of  the geometry of the Universe with the hope they will reveal us something on the nature of the agent causing the  observed accelerated expansion.

As this is quite an advanced topic in Physics (Astronomy), this contribution will build tougher as we proceed, so I hope the readers will enjoy to start off from a little historic stroll in the science of surveying the skies. For wider historical overviews than the one presented here, the two main 
sources I recommend to you, among the so many available, are \cite{conv} and \cite{wiki}.

Historical records tells us that Astronomy was born basically because of the need by agrarian societies to predict seasons and other yearly events and by  the esoteric need to place humanity in the Universe. This discipline developed in many ancient cultures (Egyptian, Chinese, Babylonian, Mayan), but it was only Greek people who cared understanding their observations and
 spreading their knowledge unlike in other cultures. Their influence was crucial as the modern scientific attitude of relying in empiricism (Aristotlean school) and
translating physical phenomena into the mathematical language (Pithagorean school) built on their way of approaching science. Unfortunately, Aristotle was far so influential that his erroneous geocentric view of the Universe was not questioned aloud until the XVIth century.

At the beginning of that century Copernicus started to spread quietly his heliocentric cosmological model, and even though he did not make much propaganda about his ideas, they deeply influenced other figures, such as Galileo, who introduced telescopes into astronomy and found solid evidence against the geocentric model. 
Another very important contribution to the subject was made by Kepler, who gave accurate characterization of the 
motions of the planets around the Sun. These findings were later on synthetically explained by Newton in his theory of gravitation, which built on Galileo's developments on dynamics (the law of inertia basically).

The XVIIIth and XIXth century brought advances in the understanding of the Universe as a whole made of many parts which lie not necessarily in the Solar system. Stars began to be regarded as far-away objects in motion, and other objects like nebulae were discovered, so the idea of the existence of complicated structures outside the Solar system gained solidity. 
A discovery very much related to the main topic of this lectures was the realization that the combination of the apparent brightness of a star and its distance get combined to give its intrinsic brightness; basically the amount of energy reaching us in the form of light from a distant source decreases is inversely proportional to the square of the distance between the source an us (in an expanding the universe definition of distance is not the same as in a static one but this broad way of speaking applies all the same) 
This finding let observers realize that stars were objects very much like the Sun, or if you prefer they realized the Sun was nothing but yet another star.

  The most important next breakthrough was  Einstein's theory of special relativity, which generalized Galileo's relativity to introduce light. On the conceptual realm, this was a very revolutionary theory at is warps the notions of space and time, and so it set the foundations for the best description of gravity so far: Einstein's theory of General Relativity.
 This second theory was the fruition of Einstein endeavors to unify the interactions known to him. In this theoretical framework matter/energy modifies the geometry, and in turn geometry tells matter/energy how to move/propagate (paraphrasing Wheeler's renowned quotation).
Among other predictions this theory made a couple which are cornerstones of modern astronomy:
 gravitational redshift (light gets redder as it moves away from massive objects), and gravitational lensing
(light gets bent as it  passes close to massive objects).

   The next important advance come from the side of observations. In 1929 Hubble, and after having collected data carefully for almost a
decade,  presented the surprising  conclusions that galaxies on average move away from us.
This effect is encoded in the scale-invariant relation known as Hubble's law: $ v=H d$, where $v$ is the galaxy's velocity and $d$ its distance from us.
The positiveness of the  quantity $H$ as measured by Hubble is precisely what told him the Universe is expanding, this being a discovery which gets accommodated nicely in Einstein's theory of general relativity.
Interestingly, on learning about this finding, Einstein discarded his idea of the necessity of  some exotic fluid with negative pressure to counteract the attractive effect of usual matter (which would make the universe contract). It cannot look but funny from today's perspective that Einstein's idea has come to life again, at the end of the day the exotic fluid he imagined is one of the possible flavors of what  the  community calls dark energy \cite{turner} these days. 

At the risk of not giving everyone the credit they merit, I will just say that recognition for the concept of a expanding Universe is due to both theoreticians (de Sitter, Friedmann, Lema\^itre, Gamow,...) and experimentalists (Slipher, Hubble,..). However, the  trampolin for this idea to jump into orthodoxy was put by Penzias and Wilson 
\cite{penzias} who first detected the cosmic microwave background, and by Dicke, Peebles, Roll and Wilkinson \cite{peebles} who were responsible for the not the less important interpretation of those observations.
The existence of this radiation and its characteristic black body spectrum are a prediction of the Big-Bang theory, it is fair to say that if one combines it with other sources of evidence, it is almost impossible to refute it.

The CMB  in an invaluable source of cosmological information (visit \cite{hu} for a higly recommended site on the subject). It has a temperature of of around $2.73\,K$, with tiny temperature differences of about $10^{-5}$ between different patches of the sky. These anisotropies inform us that the photons forming the CMB where subject to an underlying   gravitational potential which had fluctuations, and this is just and indication of density irregularities
that seeded the structure one observes today.  On the other hand, the image of the features of the CMB in different angular scales strongly favors a universe with 
no spatial curvature (flat or Euclidean on constant time slices) that is, it supports the theory of inflation. In addition, the CMB  has a saying about the fraction of the different fluids
filling the Universe.

Fortunately, the mine of cosmic surprises was far from being exhausted, and it stored a diamond of many carats in the form of a discovery which has changed greatly the mainstream view of the Universe. Up to the late 1990, no repulsive manifestation of gravity had been spotted in Nature. In 1998 astronomical measurements provided evidence that gravity can not only push, but pull as well, and the repulsive side of gravity got unveiled.
 Very refined observations of the brightness of distant supernovae \cite{Riess:1998cb,Perlmutter:1998np} seemed to hint the presence of a negative pressure component in the Universe which would make it accelerate, and then a new revolution started.

One may wonder at this stage what is the relation of supernovae luminosity and cosmic speed up.
 Objects in the Universe with a well calibrated intrinsic luminosity (supernovae, for instance) can be used to determine distances on cosmological scales. Supernovae are very bright objects, and so they result particularly attractive for this purpose (they can be $10^9$ times more luminous than the Sun so there is hope they will be visible from up to perhaps $1000$ Mpc \cite{physto}). As we anticipated in the last paragraph, in  1998 two independent teams reported evidence that some distant supernovae were fainted than expected. This involved tracing the expansion history of
the Universe by combining measurements of the recession velocity, apparent brightness and distance estimations.
The most compelling explanation was (and keeps being as far I am concerned) that their light had traveled greater distances than assumed. The orthodox view up to then was that the expansion pace of the Universe was barely constant, but supernovae seemed to contradict this. This unexpected and exciting discovery obliged researchers to broaden their mind and accept the Universe is undergoing accelerated expansion.

I should have been able to have convinced the readers by now that these are exciting times to be working on Cosmology, as cosmic speed up is such and intriguing phenomenon with major open questions such as whether dark energy evolves with time, how much of it  is there, and if is rather not a manifestation of extradimensional physics.

The answer to these questions requires cannot be dissociated from the response to a perhaps more fundamental question:
what is the Universe made of?

The combination of various astronomical observations  tell us our universe is basically made of thee major components (see for instance \cite{wmap}).
The most abundant one is dark energy \cite{turner}, so this makes it even more interesting to find out whatever we can about it. At the other end the by far least abundant component is baryonic matter, and in the middle (as abundance is concerned) we have dark matter, in a proportion comparable to that of dark energy. 
There are various sources of astrophysical giving evidence in favor of it.  Hints  of it is existence are provided by the motion
of stars, galaxies and clusters, but it is known it also played a crucial role in the amplification of the primordial density fluctuations which seeded the large scales structure we observe today and dark matter imprints 
can be found in the CMB as well. Dark matter represents quite a challenge as its nature remains a mystery; nevertheless if it were baryonic we know from big bang nucleosynthesis helium-4 would get converted to deuterium much easily
and CMB calculations indicate indicate in addition anisotropies would be much larger
so the odds are most of its is not baryonic.

Up to here we have made a very broad introduction to our topic with a little bit of history and a little bit of physics, but we must not forget Maths are key to Cosmology \cite{Tegmark}. We only know how to study the Universe  using numbers and equations, but of course progress in this direction is done with as many reasonable simplifications as possible (if they do not compromise rigor, of course).
Einstein equations relate geometry and matter/energy content of the Universe.
Cosmologists are concerned by this relation on a large scales picture so as to understand  the expansion of the Universe.
Those equations are non-linear, so studying them is painstaking unless one exploits the observational evidence the regularity of the Universe.
Non-linearity of those equations makes their study a really hard task so simplifications are a must. The two basic ones are that galaxies are homogeneously distributed on galaxies  larger than $50$ {\rm Mpc} \cite{Yadav}, and that the Universe is isotropic around us on angular scales larger than about $10$ degrees \cite{Souradeep}.
 But this is not enough, those two simplifications, which come from observational evidence must be completed with  the assumption we occupy no special place in the Universe.

This puts on the track of what we could call the parameterized Universe, cosmologists  work with a greatly a simplified geometric description of the Universe which emerges from the latter assumptions.
The models for the sources are of reduced complexity too, the most common being perfect fluids and fields with known dynamics. This we have on the side of theory, but on the side of observations we have to make our own life easier too. When doing observations oriented Cosmology, one uses  as a variable the redshift $z$
of the electromagnetic radiation received, as it encodes information of how much the Universe has expanded between emission and reception. Observational tables of geometrical quantities can be given, and then one can test different theoretical values of the same quantities 
corresponding to models of interest. The theoretical predictions will depend on the sources assumed, and  ultimately it will be possible to estimate the suitability of a given dark energy
model to observations, but which are the available  probes of dark energy?

Basically dark energy can be scrutinized observationally from two main perspectives  \cite{trotta}:
one possility is doing it through its effect of the growth of structures, another one is through its impact on 
geometrical quantities. We will concentrate on geometrical constraints/tests in these lectures as in a way they are
those which can perhaps be applied with less difficulty, although their simplicity does not mean they are the least interesting, for instance, the supernovae test is the only test giving a direct indication of the need of a repulsive component in the Universe, 
whereas the baryon acoustic oscillations test is thought to have much information in store. Discussion on these two tests will be given later on in these lectures.

Finally, there is one more direction in which these topic is related to Maths apart from the geometrical side of it, statistics plays an important role too.  Physics is attractive because of its ability to ``tame'' natural phenomena in the sense that laws of physics bring order to the apparent chaos of Nature,
Astrophysics is even more attractive because it evidences how laws of physics apply outside Earth, which is certainly surprising given the manifest differences between Earth and every other locations in the Universe of which we are aware.
However, since these phenomena occur in places far, far away, and sometimes they can only be observed indirectly, there is typically an important degree of uncertainty. Thus, research is Astrophysics requires understanding  not only Physics, but also inference.

Inferential statistics  cares for the identification of patterns in the data taking in account the randomness and uncertainties in the observations. In contrast,  descriptive statistics is concerned with giving a summary of the data either numerically or graphically. Both will be needed toward two goals: we need to know  optimal ways to extract information from the astronomical data, but we also need to know rigorous approaches to compare theoretical predictions to observations.

Our approach will be that of Bayesian inference, and we will justify jut below my preference, but it must be admitted there is an old vivid controversy on the definition of probability between the two main schools: frequentists and Bayesians.
Frequentists use a definition based on the possibility of repeating the experiment, but this does not apply to the Universe, and this sounds like the reason why the number of Bayesian astronomers grows every day, but you should not care for this battle right now; just stick to the idea that observations related Cosmology needs to resort to Statistics and that Inference is vital for constraining dark energy cosmological models. 

By now you should be able to guess what to expect in the next sections. I will present you some basics of FRW (Friedmann-Robertson-Walker) cosmologies, then I will devote a great deal of this text to details of geometrical tests, and leave for the last but one section a convenient primer on statistics.

\section{General Relativity and FRW}
 Einstein's influence in cosmology is paramount, Special Relativity is crucial in high energy physics, which plays an important role in  astronomy and cosmology, as many processes one studies in these areas are very energetic. But General Relativity is even more important for Cosmology as it allows to describe the gravitational interaction, which is the one governing the dynamics on planetary, galactic and cosmological scales. Gravity in astronomy is mostly treated classically as opposed to quantum mechanically, as no successful theory of gravity exist. The situation is different with respect to Special Relativity, as the connection  with the quantum realm is satisfactorily given by quantum field theory.
Given that Special Relativity is the conceptual precursor of General Relativity, a few lines about it are worth before entering an overview of General Relativity.

Part of the topics of this section are extensively covered in 
\cite{dinverno,tasi,goobar,peacock}.

Special Relativity stands on two pillars. The first one is that the laws of physics are the same in all inertial frames, reference systems in which bodies are not subject to forces remain at rest or in steady linear motion. The second one is that  the speed of light in vacuum, $c$, is the same in all inertial frames as confirmed by the Michelson-Morley experiment, but actually anticipated intuitively by Einstein. Inferring conclusions from those two premises requires treating the quantity $ct$
on the same grounds as 
spatial coordinates (say $x,y,z$), and this interchangeability of space and time  make the concept of spacetime emerge.
When transforming between inertial frames, the quantity $ds^2=c^2dt^2-dx^2-dy^2-dz^2$ remains invariant, $ds^2$ being the norm of the four-vector $(cdt,dx,dy,dz)$.

Special Relativity requires the laws of physics to be written in terms of four-vectors as their norms do not change on going from one inertial frame to another. 
The quantity $ds^2$ is called the line-element, and it quantifies the distance between events of the spacetime.
The line-element is a quadratic form constructed from a matrix $\mathbf g$:
$
ds^2=g_{\mu\nu}dx^{\mu}dx^{\nu}$ for $\mu,\nu=0,1,2,3$ In Special Relativity $g_{ij}={\rm diag}(1,-1,1,1)$, but in General Relativity it need not be diagonal nor constant.

This brief account on Special Relativity drives us into the theoretical ground of General Relativity.
In developing this beautiful theory Einstein was influenced by five principles: {\bf Mach's principle} (geometry or motion do not make sense in an empty universe); {\bf principle of equivalence} (the laws of physics look the same to an observer 
in non-rotating free fall in a gravitational field and to an observer in 
uniform motion in the absence of gravity), {\bf principle of general covariance} (laws of physics must have the same for all observers), {\bf correspondence principle} (from General Relativity one must recover on the one hand gravity Special Relativity when gravity is absent, and one must also recover Newtonian gravity when gravitational fields are week and motions are slow, actually, this principle reflects the very reasonable need that any new scientific framework must be consistent with precursor  reliable frameworks in their range of validity.)

According to records, walking the tortuous path from Spatial Relativity to General Relativity took Einstein 11 years. Let us outline the main elements of the construction:
gravity can be waived locally and regain Special Relativity, locally gravitational effects look like any other inertial effect, test particles are assumed to travel on geodesics (null ones if they are photons and timelike ones if they are massive), inertial forces are accounted for in the geodesic equations by terms which depend on first derivatives of the metric (which plays the role of the potentials of the theory),
gravitational fields make geodesics converge/diverge as described by some terms in the geodesic deviation equation
which depend on the Riemann tensor (which depends on second derivatives of the metric), all forms of energy act as sources for the gravitational field and this is encoded in the Einstein equations.

In General Relativity tensors play a preeminent role. The Riemann tensor or curvature tensor $R^{a}_{bcd}$ determines how geodesic deviates, and upon contraction
it gives the curvature or Ricci tensor $R{ab}=g^{cd}R_{dacb}$. Further contractions allow deriving the curvature or Ricci scalar $R=g^{ab}R_{ab}$, and finally, the connection between matter/energy and curvature is encoded in the Einstein equations
\begin{equation}
G_{ab}={8 \pi G}T_{ab}/{c^4},\end{equation}
where $G_{ab}=$ is the Einstein tensor and $T_{ab}$ is the Einstein tensor.

At this point we have now enough theoretical machinery to study the dynamics of a standard universe, but I must insist on the fact that studying the Universe without making radical but reasonable simplifications would be a intractable problem. Fortunately, progress can be made because we are lucky enough to have evidence of its regularity.
As already mentioned in the introduction, we have evidence on the one hand of the homogeneity in the distribution of galaxies on scales larger than $50$ Mpc \cite{Yadav}, 
and on the one hand  CMB experiments inform us on the isotropy around us on angular scales larger than a $10$ degrees \cite{Souradeep}. 
 If one then  invokes that our place is the Universe is not special at all, then isotropy around all its points in inferred.
Finally, there is a theorem in
 geometry which 
 tells us that if every observer sees the same picture of the Universe when looking at different directions, then the Universe is homogeneous.

These assumptions boil down into the (Friedmann-)Robertson-Walker metric. Our Universe can be viewed as an expanding, isotropic and homogeneous spacetime, and that means its line element reads
\begin{equation}
ds^2=c^2dt^2-R(t)^2(dr^2+S^2(r)(d\theta^2+sin^2(\theta)d\phi^2),
\end{equation}
with $R(t)$ an arbitrary and $S(r)$ a function which can take three distinct forms. If one defines $R(t)=a(t)R_0$, and then makes $r\to r/R_0$, a more familiar expression is obtained:
\begin{equation}ds^2=c^2dt^2-a(t)^2(dr^2+S^2(r)(d\theta^2+sin^2(\theta)d\phi^2)),
\end{equation}
where $a(t)$ is a dimensionless quantity (customarily chosen to have value $1$ at present).

The geometry of spatial sections in the FRW metric are also worth some further discussion.
The function $S(r)$ must be such the spatial sections of the RW geometry have constant curvature, so the possibilities are $S(r)=\{sin(r),r,sinh(r)\}$.
Again, a familiar form of the line-element
is obtain by making $S(r)^2dr^2\to dr^2/(1-kr^2)$, and so the three cases are respectively
$k={1,0,-1}$ (as shown in Fig. 1).
\begin{figure}[htbp!]
\includegraphics[width=0.3\textwidth]{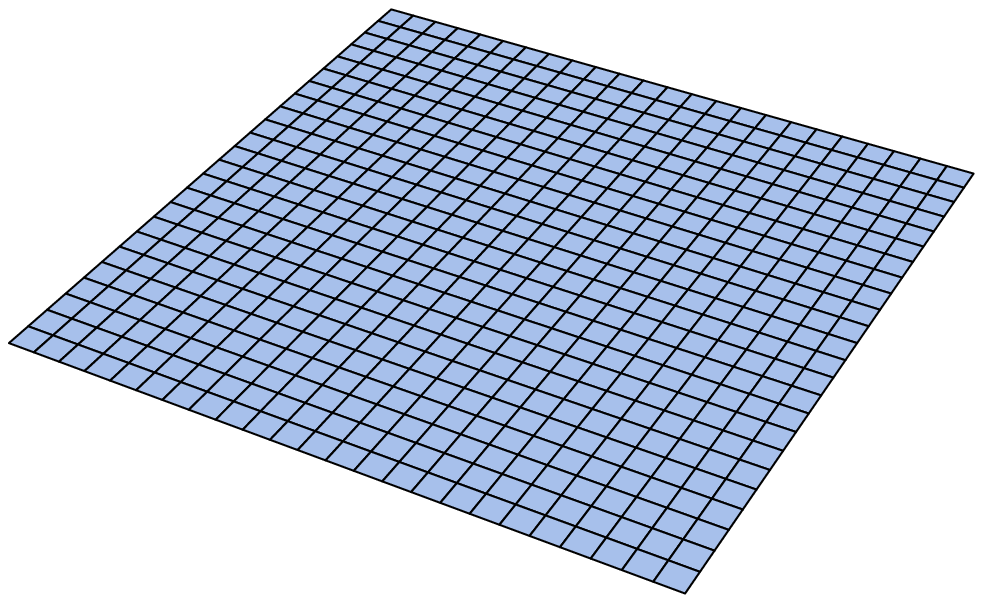}
\includegraphics[width=0.3\textwidth]{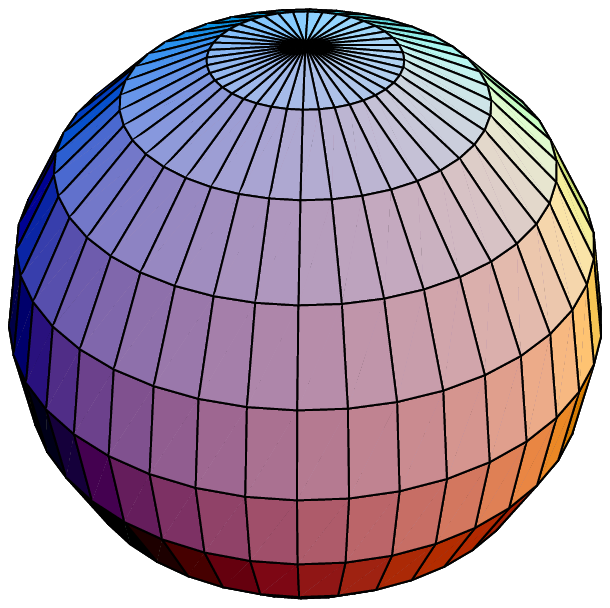}
\includegraphics[width=0.3\textwidth]{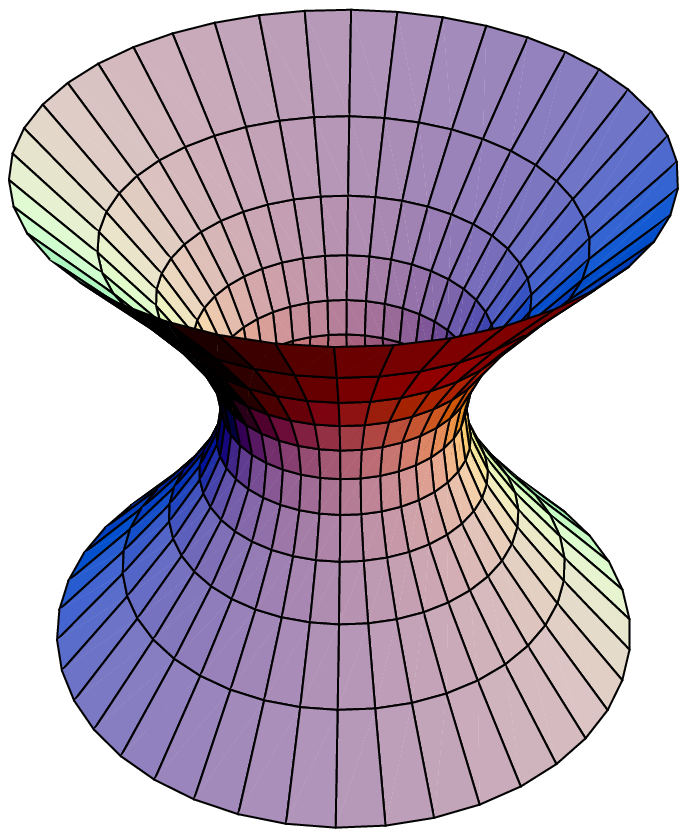}
\caption{From left to right, geometry of spatial sections in a flat universe (k=0), a positively curved or closed universe  (k=-1), and negatively curved or open universe  (k=-1)}
\end{figure}

Before proceeding, a little remark is convenient.
From here on we will use the so called natural units $8\pi  G=c=1$ under otherwise stated.
Having made this comment in passing let us turn our attention to the right hand side of the Einstein equations, i.e to the matter/energy content.

We have seen that simplifications in geometry are required to model the Universe.
In the same spirit reduction of sophistication in the description of matter/energy is also required. 
Simplicity on the one hand, and consistency with observations in the other, suggest adopting the perfect fluid picture (no viscosity is assumed):
\begin{equation}
T_{ab}=(\rho+p)u_au_b+pg_{ab},
\end{equation}
with $\rho$, $p$ and $u_a$ representing the energy density, pressure and velocity field of the fluid. Note that in the rest frame of the fluid
\begin{equation}
T_{ab}={\rm diag}(\rho,-p,-p-p).
\end{equation}

Let us now formulate Einstein equations for perfect fluids.
 There are basically two of them in these cases. The first one is the Friedmann equation,
\begin{equation}
H^2\equiv{\dot a^2}/{a^2}=\sum_{\tiny i}\rho_i/3-{k}/{a^2},\end{equation}
and it acts as a constraint as $\dot a$ is not free,  but is rather subject to the amount of energy density and curvature. The second one is the Raychaudhuri equation:
\begin{equation}
2({\ddot a}/{a}-{\dot a^2}/{a^2})=-\sum_{\tiny i}(\rho_i+p_i)+{k}/{a^2},
\end{equation}
and it is an evolution equation. The combination of the two equations can be used to derive $a(t)$ (or $t(a)$ in the least fortunate cases), once $\rho$ and $p$ have been specified.

From Einstein equations one can derive other two important equations.
The first one is the energy conservation equation
\begin{equation}
\dot\rho_{\rm tot}+3H(\rho_{\rm tot}+p_{\rm tot})=0 \qquad \rho_t=\sum_{\tiny i}\rho_i\quad p_t=\sum_{\tiny i}p_i.
\end{equation}
The second one is the acceleration equation, which tells us about the evolution of the spatial separation between geodesics
\begin{equation}2{\ddot a}/{a}=-(\rho+3p)/3.\end{equation}

These preliminaries suggest the interplay between the matter/energy content of the Universe and its geometry
have a crucial influence in its final fate.
If we consider a model with matter only then $\rho_{\rm tot}=\rho_m=\rho_0/a(t)^3$, and thus
\begin{equation}
\dot a^2 ={\rho_0}/{3a(t)}-k,
\end{equation}
so, any value of $a(t)$ is consistent for $k=0,-1$ for not for $k=1$, there is an upper bound to $a(t)$.
When $a(t)\equiv a_{crit}={\rho_0}/3$ the model will begin to collapse (because $\ddot a\ge-a/2$, which signals a maximum in the $a$ versus $t$ plot).
In contrast,  open or flat universes do not experience any particular behavior when the critical density is reached.

The addition of a cosmological constant brings a richer set of possibilities (see Fig. 2(a)).
It is convenient to present the cases using fractional densities:
\begin{equation}\Omega_k=-{k}/{H^2a^2}, \quad\Omega_{\Lambda}={\Lambda}/{3H^2}, \quad\Omega_m={\rho_m}/{3H^2},\quad
\Omega_m+\Omega_{\Lambda}+\Omega_k=1.
\end{equation}
\begin{figure}
\begin{tabular}{c@{\hspace{2cm}}c}
\includegraphics[width=0.3\textwidth]{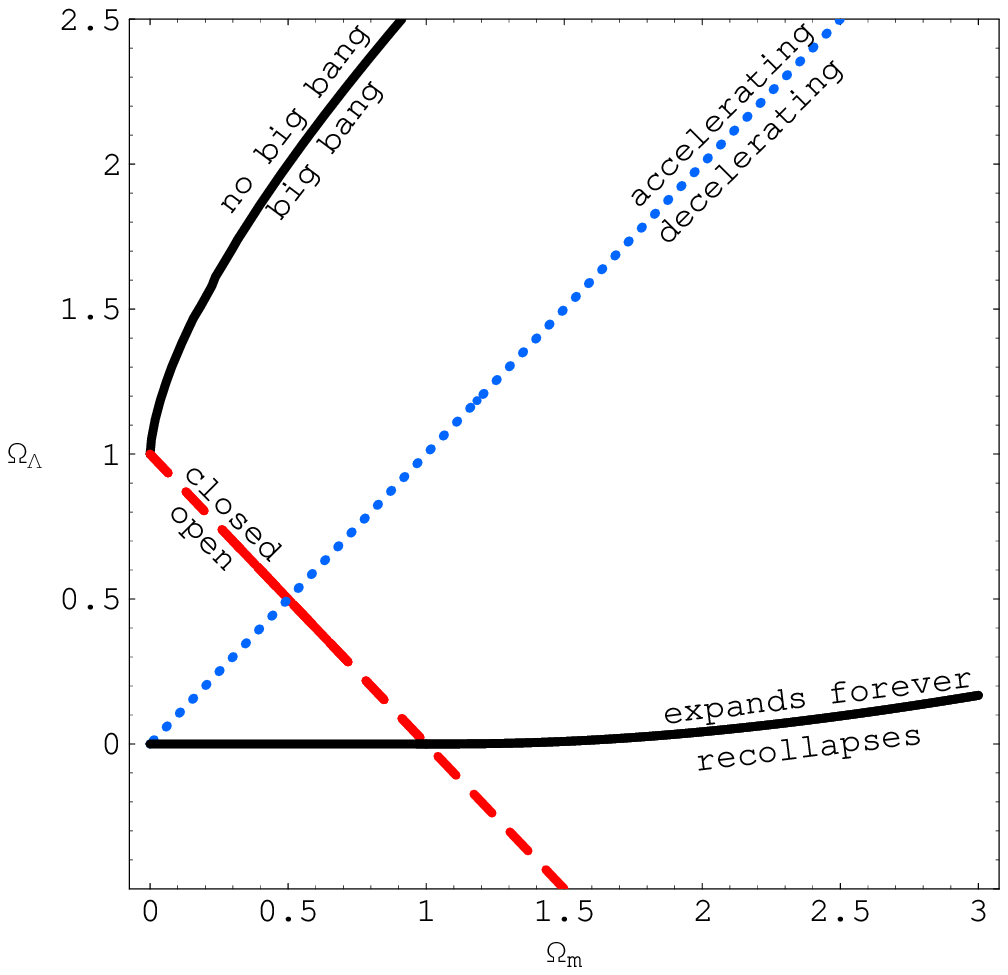}&\includegraphics[width=0.3\textwidth]{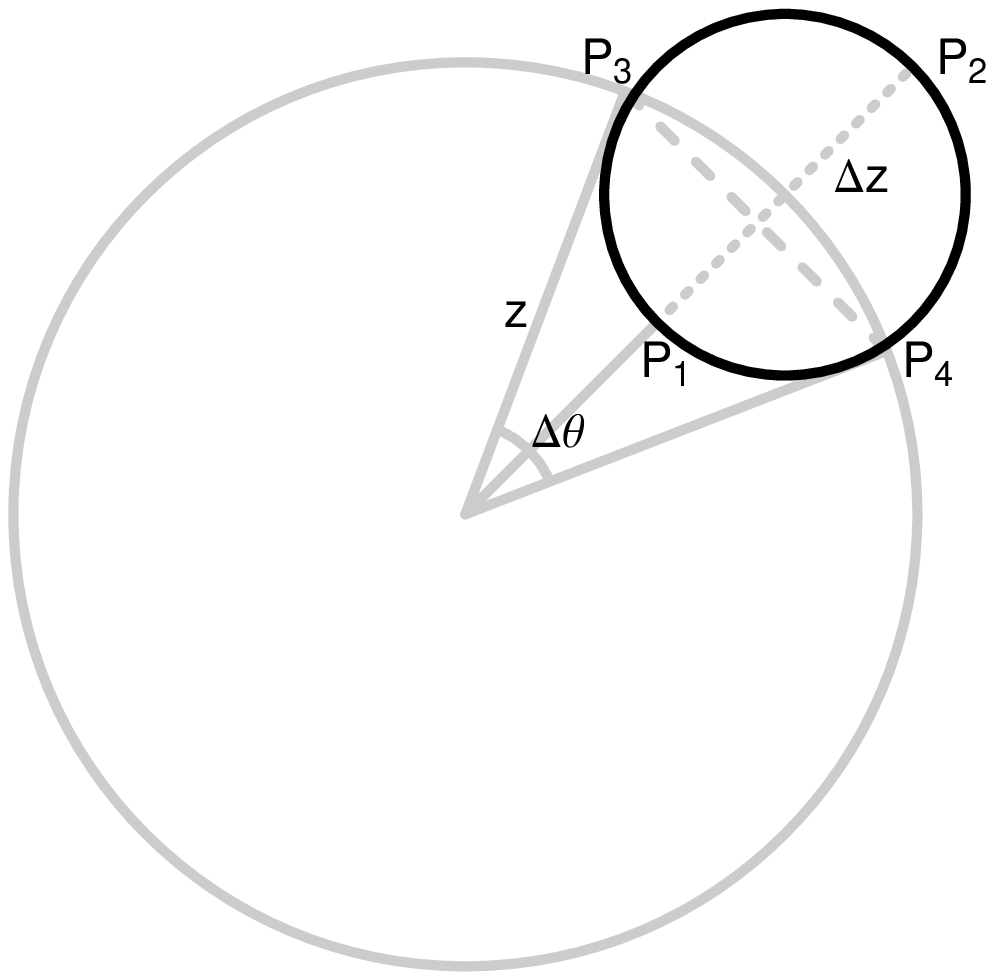}\\
(a)&(b)
\end{tabular}
\caption{(a)  Regions for the LCDM model. (b) A spherical shell in redshift space.}
\end{figure}
Let us consider now perfect fluid cosmological histories.
 Many $p(\rho)$ equations of state are considered in the literature, but the linear one, $p=w\rho$,
stars in popularity. Here you have a list some physically meaningful cases in the $k=0$ case.
\begin{itemize}
\item Electromagnetic radiation, i.e. photons,
$w={1}/{3}, \qquad \rho\sim{a^{-4}},\quad a(t)\sim t^{1/2}$
\item (Incoherent) matter, aka cosmic dust,
$w=0,\qquad\rho\sim{a^{-3}},\quad a(t)\sim t^{2/3}$
\item Vacuum energy,  aka cosmological constant,
$w=-1,\qquad\rho\sim cons, \quad a(t)\sim e^{Ht}$
\item Quiessence \cite{quiess},
$w=cons<-{1}/{3}\neq -1, \quad \rho\sim{a^{-3(1+w)}},\quad a(t)\sim t^{2/(3(1+w))}$
\end{itemize}

 There are, of course, more complicated equations of state that have received attention in connection with 
late-time acceleration. A list (with some bias, as that of perfect fluids) can be this one:
\begin{itemize}
\item Conventional Chaplygin gas \cite{Gorini:2002kf},
$\quad p=-{A}/{\rho}, \quad \rho\sim\sqrt{A+{B}/{a^6}}$
\item Generalized Chaplygin gas \cite{genchap},
$\quad p=-{A}/{\rho^{\alpha}},\quad  \rho\sim\left(A+{B}/{a^{3(1+\alpha)}}\right)^{1/(1+\alpha)}$
\item Inhomogeneous equation of state \cite{inhomo}, $p=-\rho-A\rho^{\alpha},\; \rho^{1-\alpha}\sim A(1-\alpha)\log a$
\end{itemize}

Last but not least, another very popular class of sources in Cosmology is that of scalar fields.
Usually, they can be interpreted as perfect fluids, but the main difference is that the equation of state changes over time. Many scalar field models have been proposed (for the early and the late universe):
\begin{itemize}
\item quintessence \cite{quintessence}, $\,\rho=\displaystyle{\dot\phi^2}/{2}+V(\phi)\qquad p=\displaystyle{\dot\phi^2}/{2}-V(\phi)$
\item k-essence \cite{k-essence}, $\rho=V(\phi)\bigg(2\dot\phi^2 \displaystyle{\partial F(-\dot\phi^2)}/{\partial \dot\phi^2}-F(-\dot\phi^2)\bigg)\quad p=V(\phi)F(-\dot\phi^2)$
\item tachyon \cite{sen}, $\,\rho={V(\phi)}/{\sqrt{1-\dot\phi^2}}\quad p=-V(\phi){\sqrt{1-\dot\phi^2}}$
\item quintom \cite{quintom}, $\,\rho=({\dot\phi_1^2}-{\dot\phi_2^2})/{2}+V(\phi_1,\phi_2)\quad p=({\dot\phi_1^2}-{\dot\phi_2^2})/{2}-V(\phi_1,\phi_2)$
\end{itemize}

This schematic account of the possible types of matter/energy content in Cosmology does not make justice at all to the vast literature on the subject, but we must stop here and return to geometric aspects.

Commonly, cosmological parameters are constrained by studying how distant light sources are seen from our detection devices, and this depends on the cosmological model. In an expanding FRW model light gets redshifted during its trip from the source to the observer, and the redshift depends on the amount of expansion occurred meanwhile:
\begin{equation}{a(t_{obs})}/{a(t_{emit})}=1+z\equiv {\lambda_{obs}}/{\lambda_{emit}}.\end{equation}
As light travels along null geodesics, one can compute the redshift experienced by a photon as it travels 
a given radial distance
\begin{equation}dt^2-a^2dr^2=0 \,{\rm \;implies\;}\, dr={dz}/{H(z)},\end{equation} we have taken $a_0=1$ and $c=1$ in $dr={c dz}/{H(z)}$ and the metric is the FRW one.
Those are basic ingredients for the construction of ``distances'' in cosmology (see \cite{Hogg} for a pedagogical acount of the topic).

Now, there is a basic concept we need so as to make progress in the topic of distances in Cosmology: comoving coordinates. One of the niceties of General Relativity is it allows to formulate physical laws in whatever system of coordinates one may prefer. In Cosmology, those coordinates  have become very popular because they make one's life a lot easier.
Observers which see the Universe as isotropic have constant values of spatial coordinates when the comoving system is used. Non-comoving observers will see redshifts in some directions and blueshifts in others, and
 the time measured by comoving observers is cosmological time.

One of the quantities required for our geometrical tests is the line of sight comoving distance.
 This is the comoving separation between two objects with the same angular location but different radial position ($P_1P_2$ in the picture). The comoving distance from us to an object at $z$
\begin{equation}
D_C=\int_0^z{dz}/{H(z)},\end{equation}
whereas from the former we can derive that the comoving distance between two objects separated $\Delta z$ is
$D_C={\Delta z}/{H(z)}$. The comoving distance is the proper distance between those objects divided by the ratio of the scale factor of the Universe at the epochs of emission of reception.

Another relevant definition is that of (transverse) comoving distance and angular diameter distance.
The comoving  distance between two objects located at the same radial position but separated by and angle $\Delta \theta$ is $D_M\Delta \theta$ ($P_3P_4$ in Fig. 2(b))  where 
$D_M=S(r)$ 
is the transverse comoving distance. Closely related to the former we have the widely used
 angular diameter distance $D_A$. It is defined as the ratio of an object's physical transverse size to its angular size (in radians):
 \begin{equation}
D_A={D_M}/{1+z}.
 \end{equation}

Finally, let me present the luminosity distance which is the key to the extraction of information from supernovae data.
 Given a standard candle its bolometric (i.e. integrated over all frequencies)
emitting power right at the position of the source is called the luminosity $L$.
The total bolometric power per unit area at the detector is called the flux $F$. 
The quantities $L$ and $F$ are used to define the luminosity distance 
$D_L=\sqrt{L/4\pi F}.$
At the moment of detection photons are passing through a sphere of proper surface area $4\pi (a(t_{obs})S(r))^2$.
The flux is affected by redshift in two ways: the energy decreases by a factor $(1+z)$ and the arrival dates 
are reduced by a factor $(1+z)$ so the flux is $(1+z)^2$ times smaller than in a static universe, so finally
\begin{equation}D_L=(1+z)\int_0^z{dz}/{H(z)}.\end{equation}
Actually, there is yet one more definition of distance which is used in one of the tests to be discussed below, but I prefer to postpone its mention for now. 

\section{Supernova as dark energy probes}

Theory has played the preeminent role in the development of cosmology till just a few years ago, with Einsteins' theory of General Relativity being the most compelling framework for the study of the evolution and fate of the Universe on large scale (we exclude the origin of the Universe from the list as the necessity to account for quantum effects makes General Relativity insufficient). However the preminence of theory seems doomed,  as of recent the situation has changed greatly with the rise of advanced technological resources.
 Observations allow backing up predictions of General Relativity precisely and also make researchers formulate new questions.

 Daily experience connects us with the attractive side of gravity. It keeps us attached to the ground, it allows for the fun in games involving balls, and it holds
artificial satellites revolving around the Earth.
Yet gravity has a repulsive side which fits in Einstein's theory of gravity,
but it was for decades  believed to be just a theoretical possibility. No wonder, though, as it manifests on cosmological scales only. 
The existence of a exotic component in the cosmic budget which makes the Universe accelerate was inferred from the observation of
distant supernovae
\cite{Riess:1998cb,Perlmutter:1998np}

In order to understand the evidence found in those experiments and why it was so important, some background material is required.  Supernovae are spectacularly luminous objects arising from explosions of massive supergiant stars.
 In the explosion a lot or all of the star's material is spelled out at a velocity of up to a tenth the speed of light.
These phenomena have long intrigued astronomers, for instance Chinese records date back to AD 185, whereas first european record dates back to AD 1006 \cite{suprec}.
Supernovae a are classified according to the the shape of their light curves and the nature of their spectra
\cite{Filippenko:1997ub}, and
they are fantastic for Cosmology as they may shine with the brightness of one thousand million suns and release a total of $10^{44}$ joules (you could use it to provide the average USA consumption
for $10^{10}$ trillion years).  
Unfortunately supernovae are rare, in a given galaxy they only occur twice every thousand years, so there not as many of them available as cosmologists would wish.

Let us now present some basic facts about supernovae.
 Type Ia supernovae (SNe Ia) are typically 6 times brighter than other supernovae, so predictably those are the most frequent supernovae in high redshift surveys. These explosions occur in binary star systems formed by a carbon-oxygen white dwarf and a companion star. The white dwarf accretes mass from the companion and when the Chandrasekhar limit is reached ($10^{14}$ solar masses), the nucleus gets fused suddenly and the star explodes getting completely disrupted.

 An important part of the game are the different tools to determine the distance to a given supernovae (or to an astronomical object in general). The  magnitude of a star is defined through its flux as $m=-2.5\log_{10} F+const,$so the brighter the star the lower the magnitude (there are plenty of places where this is explained, one is 
 \cite{mags}. The photometric zero point  
\cite{HST}
 is related to Vega
so the magnitude difference between two stars is  
\begin{equation}m_1-m_2=-2.5\log_{10} F_1/F_2.\end{equation}
The mnemotecnic rule is that a flux ratio of 100 gives a magnitude difference of 5.
As flux $F$ is related to luminosity distance through $F=L/4 \pi D_L^2$, for two objects of the same intrinsic luminosity 
$m_1-m_2=5\log_{10} D_{L1}/D_{L2}$.
The absolute magnitude of a star is denoted by $M$ and is defined as its  apparent magnitude at  a distance of 10 parsecs (32.6 light years) 
so
$m-M=5 \log_{10}(D_L/10 pc)$
and the quantity $m-M$ is called the  distance modulus.

Another important aspect of the problem is to what extent  supernovae can be considered as standard candles.
 SNe Ia are not truly standard candles as they do not display the same luminosity at maximum 
and their appearance is not uniform. However,  on the safe side we can say supernovae are standarizable as they can be brought in line with each other by some corrections: these are the ``stretch factor correction'' \cite{stretch}
and the `K-correction'' \cite{kcorrect}, 
which are respectively a stretching or a contraction of the timescale of the event and 
a correction to compensate for the slight differences in the part of the spectrum observed by  filters used to observe high and low redshift supernovae.

Now let us enter the core of this section which is how to set up constraints on the parameters of the Universe using supernovae. As we have just mentioned, supernovae are not standard candles, but their luminosity curves can be unified to obtain a template value of the absolute magnitude. Photometry gives us the apparent magnitude $m$ and our massaging of the light curves allows fixing $M$.
This is only half of the story, though, because we also need to
associate a recession velocity (or redshift $z$ to each supernova),
get rid of nuisance parameters (if possible), and formulate one's preferred model and test it.
These were basically the steps followed in the pioneering works of 1998, and these are (more or less) the same steps everyone else in this factory keeps doing.

In what 
recession velocities are concerned it must be kept in mind that
astronomical objects have characteristic spectral features due to the emission of specific wavelengths of light from atoms or molecules. Templates exist which can be used to determine which kind of object one is observing one basically compares different spectra till a best match is found
and which is  the object's redshift.
Supernovae are assigned the redshift of their host galaxy; Fig. 3 (a) 
  is a courtesy of SDSS \footnote{\tiny Image courtesy of SDSS. 
Funding for the SDSS and SDSS-II has been provided by the Alfred P. Sloan Foundation, the Participating Institutions, the National Science Foundation, the U.S. Department of Energy, the National Aeronautics and Space Administration, the Japanese Monbukagakusho, the Max Planck Society, and the Higher Education Funding Council for England. The SDSS Web Site is http://www.sdss.org/.
The SDSS is managed by the Astrophysical Research Consortium for the Participating Institutions. The Participating Institutions are the American Museum of Natural History, Astrophysical Institute Potsdam, University of Basel, University of Cambridge, Case Western Reserve University, University of Chicago, Drexel University, Fermilab, the Institute for Advanced Study, the Japan Participation Group, Johns Hopkins University, the Joint Institute for Nuclear Astrophysics, the Kavli Institute for Particle Astrophysics and Cosmology, the Korean Scientist Group, the Chinese Academy of Sciences (LAMOST), Los Alamos National Laboratory, the Max-Planck-Institute for Astronomy (MPIA), the Max-Planck-Institute for Astrophysics (MPA), New Mexico State University, Ohio State University, University of Pittsburgh, University of Portsmouth, Princeton University, the United States Naval Observatory, and the University of Washington.}.

\begin{figure}
\begin{tabular}{c@{\hspace{2cm}}c}
\includegraphics[width=0.4\textwidth]{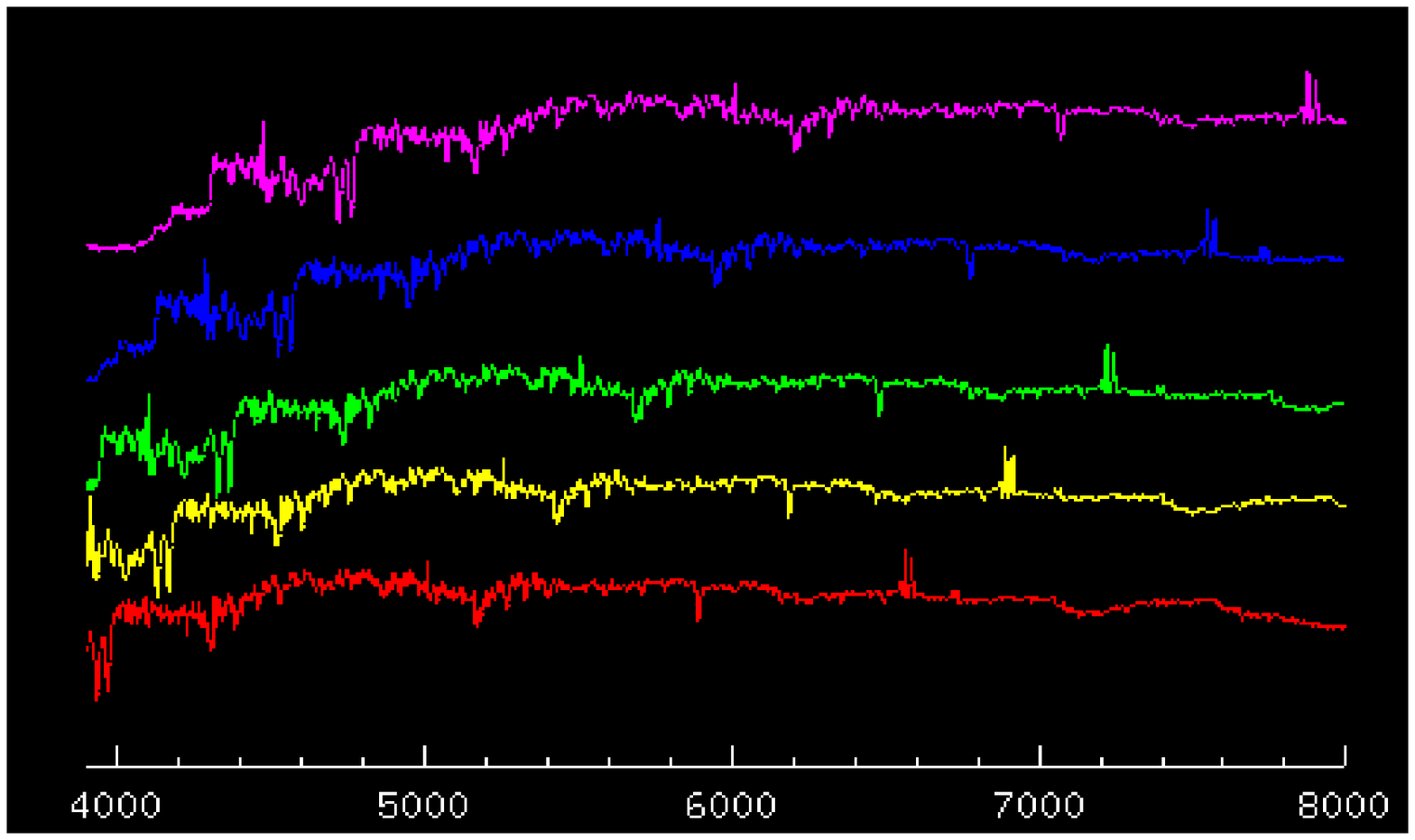}&\includegraphics[width=0.4\textwidth]{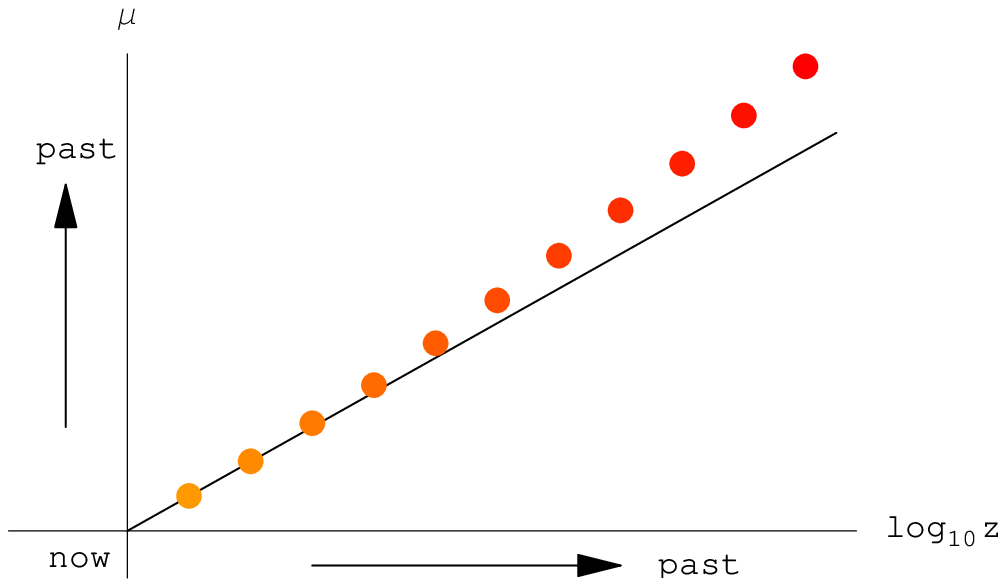}\\
(a)&(b)
\end{tabular}
\caption{(a) A galaxy spectrum at four different redshifts (0.0, 0.05, 0.10, 0.15, 0.20). (b) Mock Hubble diagram}
\end{figure}
 
The trick now is to draw the Hubble diagram of your supernovae sample(s) (see Fig. 3(b)). Recall that the larger $m$ the fainter the object.
 At low redshifts ($z\leq 0.1$) one has a linear law
$m-M=\mu\simeq 5 log_{10} z+\mu_0$. But at high redshifts it is no longer  valid and
$m-M=\mu\simeq 5 log_{10} d_L(z)+\mu_0$
and the role of  the matter and dark energy content become substantial.
As one can deduce $d_L(z)=z+(1-q_0)z^2/2+\dots...$, and  observations indicate $d_L(z)>z$ at high redshifts, 
this is evidence indicating  $q_0<0$ i.e. a currently accelerating universe. As  
$q_0$ depends on the cosmological model (e.g. $q_0=\Omega_m/2-\Omega_{\Lambda}$ in LCDM) 
it gives an observational way to constrain the universe. On the other hand, by definition
$d_L(z)=(1+z)\int_0^z H_0/H(z)$, from where it follows that what the observations are telling us is that  $H(z)>H_0$ in the past, which is another way of saying the rate of expansion has increased.  

Those best fits are obtained by minimizing the quantity
\begin{equation}\chi^2_{\rm SN}(\mu_0,\{\theta_i\})=\sum_{j=1}^{N}{(\mu_{\rm th}(z_j;\mu_0,\{\theta_i\})-\mu_{\rm obs}(z_j))^2}/{\sigma_{\mu,j}^2}, \end{equation}
the $\sigma_{\mu,j}$ being the measurement variances \cite{Davis}.
The nuisance (statistically non-important) 
parameter $\mu_0$ encodes the Hubble parameter and the absolute magnitude $M$ 
and has to be marginalized over
\cite{rob}, so one will
actually use be working with the quantity
$$
 \hat\chi^2_{\rm SN}({\{\theta_i\}})=-2\log\left(\int e^{-\chi^2_{SN}(\mu_0,\theta_1,\dots,\theta_n))}
d\mu_0\right).$$
 These $\chi$ and $\sigma$ quantities deserve an explanation but I will postpone it for a few sections yet.
 
Finally, in case you want to undertake studies of this sort by yourself you may like to know that at the time
of writing this contribution  the Davis et al. 2007 dataset \cite{Davis} is one of the latest supernovae catalogs to be completed.
It consists of 192
SNe classified as type Ia up to a redshift of $z=1.755$. It is formed by 60 ESSENCE supernovae 
\cite{Wood-Wasey}, 
57 Supernova Legacy Survey supernovae \cite{Astier},
45 nearby supernovae 
\cite{Hamuy,Riess98,Jha} 
and 30  Hubble Space Telescope (HST) supernovae \cite{Riess07};
and it is available at 
\cite{davisset}
Now, even Supernovae luminosities give the strongest  evidence of the current acceleration of the universe.
(and letting alone the problem with the physics of supernovae themselves), there is the problem  
of a certain degeneracy in the test (see Fig. 4) so supernovae do not provide a good individual estimate of the cosmological parameters. We will address some of those tests in the next sections.

\begin{figure}
\includegraphics[width=0.4\textwidth]{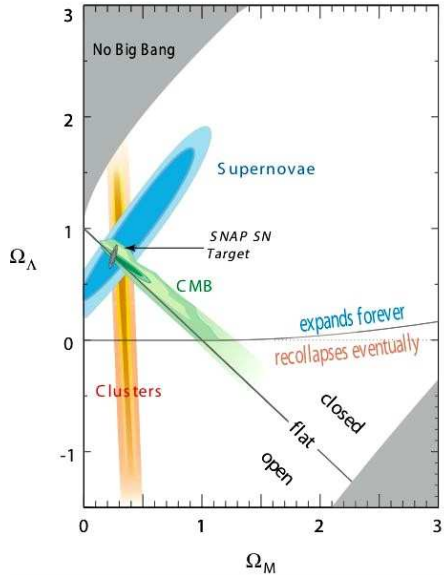}
\caption{68$\%$ and 95$\%$ confidence regions of $\Omega_m$ and $\Omega_{\Lambda}$ from
 various current measurements, and expected confidence region from the SNAP supernova program.  Image courtesy of SNAP.}
\end{figure}

\section{Hubble parameter from stellar ages}
There is a feverish activity in constraining parameters of the Universe. We are focusing in these lectures in the tests based solely on geometry. The first goal of these researches is determining the current value of the equation of state parameter $w$.
The next goal is to determine its evolution, i.e. to draw its redshift history.
The supernovae luminosity test is somewhat compromised by its integral nature:
$d_L(z)=H_0(1+z)\int_0^z(1+z')({dt}/{dz'})$.
As an alternative, measures of the integrand of the latter have been proposed in
\cite{Jimenez:2001gg} as a way to improve sensitivity to $w(z)$.
That test relies on the availability of a cosmic clock to measure how the age of the Universe varies with redshift.
Such a clock is provided by spectroscopy dating of galaxy ages.
The idea is to obtain the age difference $\Delta t$ between two-passively evolving galaxies born approximately at the same time but with a small redshift separation $\Delta z$.
One will then use $\Delta t/\Delta z$ to infer $dt/dz$, as it is directly related to the Hubble parameter
$H(z)=-{(1+z)}^{-1}{dz}/{dt} \label{H(z)}$,
which is just the inverse of the integrand in the luminosity distance formula.

However, in Astrophysics important open questions about galaxy formation and evolution remain.
How did a homogeneous universe become a clumpy one? Inflation drives a major early homogeneization of the Universe but it also provides the primordial fluctuations which give rise to structure.
How were galaxies born? How do they change over time? Progress in understanding these questions will help making the most out of this test.
 
 Hierarchically superclusters, clusters, galaxies, star clusters and stars can be distinguished in what we are concerned in this section. 
Clusters may contain thousands of galaxies, and are often associated in larger groups 
called superclusters.
Globular clusters are spherical collections of stars orbiting around a galactic core (so they are its satellites)
are very tightly bound by gravity, and the oldest of them contain the oldest stellar populations known. They have been estimated to have formed at $z>6$. It is these objects one resorts to in order to determine $H(z)$.

 \begin{table}[hbtp!]\caption{H(z) data  (in units of
kms$^{-1}$Mpc $^{-1}$)}
\begin{tabular}{l|@{\hspace{0.2in}}l@{\hspace{0.2in}}l@{\hspace{0.2in}}l@{\hspace{0.2in}}l@{\hspace{0.2in}}
l@{\hspace{0.2in}}l@{\hspace{0.2in}}l@{\hspace{0.2in}}l@{\hspace{0.2in}}l@{\hspace{0.2in}}}
\hline
z &0.09& 0.17& 0.27&0.4 &0.88 &1.3 &1.43 &1.53& 1.75\\ 
H(z) &69 & 83 &  70 & 87 & 117 & 168 & 177 &  140 & 202 \\
$\sigma$&  12 &8.3 &14& 17.4 &23.4& 13.4&14.2&14 &40.4\\
\hline
\end{tabular}\label{Htable}
\end{table}

On the other hand, most galaxies in cluster are elliptical galaxies, and of course this stands for the oldest galaxies in clusters too. This allows for some simplifications in their spectroscopic analysis.
The derivative of cosmic time with respect to redshift is inferred from the aging of stellar populations in galaxies.
Even though birth rate of galaxies is very high at high redshifts, examples of passively evolving (typically old and red)  galaxies are known.   

In \cite{Simon:2004tf}
three such samples were used: the first one was a sample of old red galaxies from the Gemini Deep Deep Survey, the second one was the so called Treu sample (made of field early-type) galaxies, and finally they used two radio galaxies (53W091 and 53W069). A collection of 32 galaxies was completed this way and their ages were estimated using SPEED stellar population models \cite{speed} which were confronted with estimates by the GDDS collaboration \cite{gdds} and were shown to give a good agreement.

The next task was deriving differential ages, and this was done in  various steps. Firstly, all galaxies within $\Delta z=0.03$ of each other were grouped together.  This allowed estimating the age of the Universe at a given redshift. The redshift interval is small enough to exclude galaxies which have undergone a significant age evolution, but large enough as for the bins to be made of more than one galaxy. 
Secondly, age differences were calculated by comparing bins with $0.1<\Delta z<0.15$.
The lower bound results in an age evolution larger that the error in age determination;
Thus, one achieves a robust age determination.
 Finally, the value of $H(z)$ was derived using the convenient expression given above, and then  Table \ref{Htable} was constructed.

\section{Cosmic microwave background}
The first announcement of the discovery evidence of the cosmic microwave background (CMB) dates back 42 years.
 The radio astronomers Penzias and Wilson did the the discovery but did not realize what it was due to. In broad terms they detected an extraterrestial isotropic excess noise using an Bell Labs antenna sited at Holmdel (New Jersey) which after years of being used for telecommunications had by 1962 been freed up for pure research
Dicke's group at Princeton (formed by Peebles, Roll, Wilkinson and Dicke himself) gave the theoretical interpretation of the Penzias and Wilson result as a prediction of the big bang models.
Two papers \cite{penzias,peebles} were sent jointly and appeared in the 142th volume of the Astrophysical Journal, the Penzias and Wilson paper was humbler, but they got rewarded years later (in 1978) with the Noble Prize.
Theoretical advances  and predictions on the topic had been done earlier by Gamow, Alpher and Herman, and although the Princeton group had drawn their conclusions independently, one of his members, Peebles, recognized the contribution of those authors in his classical textbook \cite{peeblesb}. 
Just a few months after the publication of the mentioned two papers, Roll and Wilkinson \cite{roll} confirmed in another work the thermal nature of the spectrum of the radiation 
(in concordance with the big bang model prediction) 
and estimated its temperature to be about $3.0\pm 0.5 K$.

The next major breakthrough in this topic was the discovery of the CMB anisotropies by the COBE satellite, and as a consequence
Cosmology got two more Nobel laureates in 2006: Smooth and Mather. 
According to the calculations carried out  by Harrison \cite{Har}, Peebles and Yu \cite{Pee}, and independently by Zeldovich \cite{Zel},
CMB anisotropies  are the consequence of  the predictable primeval inhomegeneities of amplitude $10^{-5}$. 

The CMB power spectrum, through its pattern of peaks (see Fig. 5(a)),
tell us that the Universe is very close to spatially 
flat (at a high degree of accuracy) (see for instance \cite{sarah}
among the also vast amount of references mentioning this).
The CMB also inform us that inflation was the major component of cosmic structure formation as opposed to cosmic strings (for a short history of the discovery of the first peak see \cite{firstpeak})

You may already be convinced that CMB physics is capital.
On the one hand no other cosmological probe has beaten it so far in the amount of relevant information it provides, and on the other hand it supports the big bang model and links Cosmology and particle physics 
as the observed abundances of light elements determined from measurements of the fluctuations of the cosmic background radiation temperature are in agreement with big bang nucleosynthesis \cite{Schramm}, being based on the well-trusted Standard Model of Particle physics. But the list of important pieces of information it gives us does not end there as it provides traces of very weak 
primordial fluctuations which may have seeded the large scale structure observed today, it informs us of the geometry of the universe (first peak), it provides evidence for baryonic dark matter (second peak), it constrains the amount of dark matter (third peak) \cite{peaksinfo}
and finally, a fact which is very important to us is that  constraints on dark energy models can be refined using CMB data.

\begin{figure}
\begin{tabular}{c@{\hspace{2cm}}c}
\includegraphics[width=0.4\textwidth]{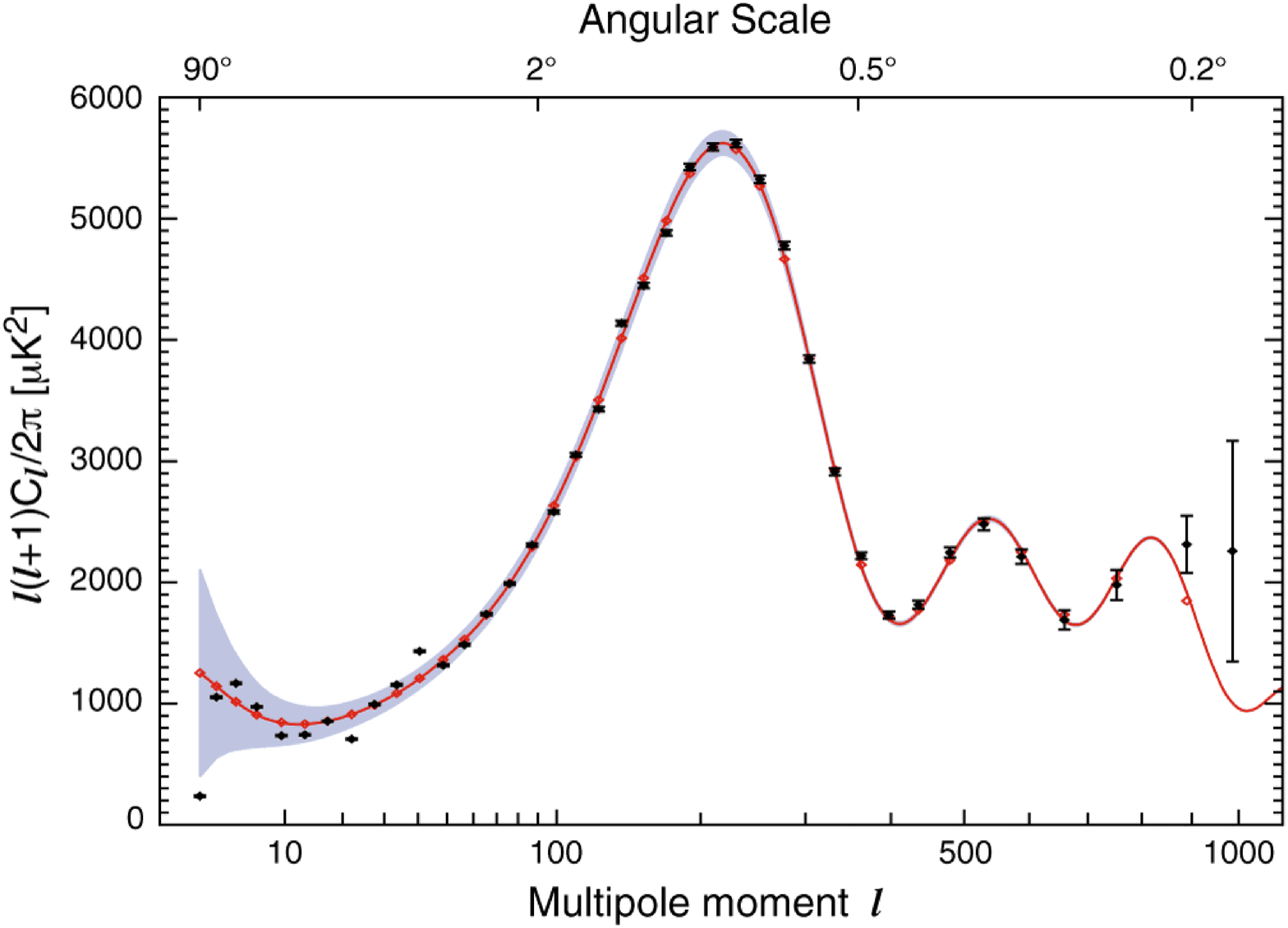}&\includegraphics[width=0.4\textwidth]{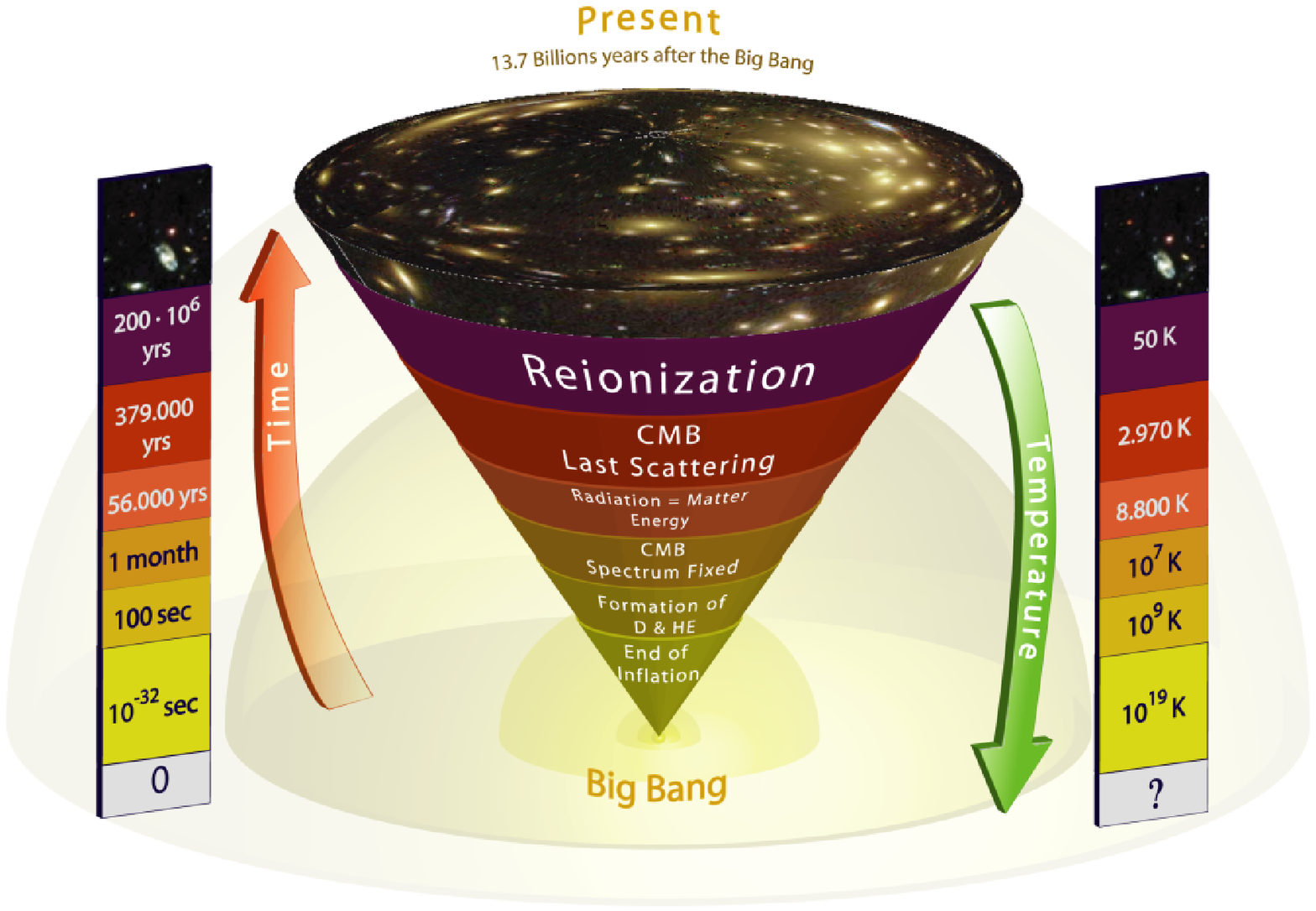}\\
(a)&(b)
\end{tabular}
\caption{(a) WMAP angular temperature power-spectrum. (b) Timeline of the universe. Image kindly produced by Ra\'ul B. P\'erez-S\'aez.}
\end{figure}

At this stage is it convenient to make a thermal history sketch (as illustrated in Fig. 5(b)). After the big bang the Universe cools down due to adiabatic expansion and its goes through
these stages: the energies of particles decrease and several phase transitions occur,
massive particles become non-relativistic as temperature decreases below their rest mass, the rates of production of particle-antiparticle decrease and anhilitation leaves asymmetric populations, nuclei of light elements form  ($z\simeq 10^4$) and the Universe becomes a soup of nucleons, photons and free electrons (the CMB forms at $z\simeq 10^7$).

As the density of relativistic particles decreases faster (proportional to $a^{-4}$) 
than that of non-relativistic ones (proportional to $a^{-3}$) 
at some stage one gets matter-radiation equality.
Later on, thermal ionization ceases to be effective, atoms form ($z\simeq 10^4$)  electrons get bound to nucleons (recombination)
and photons decouple and travel freely to us since
\cite{thermal1,thermal2}


Just to give a few more details about the formation of the CMB background let us mention that
it formed during the epoch $10^{10}>z>10^7$ due to photon creation processes (bremsstrahlung  and double Compton scattering) \cite{Birkel:1997be}. 
At later stages the dominant coupling effect turned to be Compton/Thompson scattering but it did not destroy the thermal nature of the spectrum. 

The basic observable in CMB physics is temperature fluctuation $\Delta T/T$  (a field depending on direction on the sky).
 Initial density fluctuations were tiny, with an amplitude
 of about $10^{-5}$ so their evolution is linear, i.e. Fourier modes evolve independently.
 These primordial irregularities give rise to potential wells and hills according to which the baryons and the photons
 get accommodated.
Photons,  climb up (down) potential wells (hills) in an unimpeded trip to us after getting released at the decoupling redshift $z_{dec}\approx 1100$ and 
give us  a snap-shot of the early universe in the form of cold and red spots.
This is a primary anisotropy that gets imprinted on the CMB at last scattering
responsible for the characteristic large scale anisotropy called
Sachs-Wolfe plateau \cite{Lineweaver:1996tk} (this anisotropy is associated with fluctuations
with period so large they have had no time to perform a complete oscillation by the recombination
time \cite{rosetta}).

Let us now discuss were the peaks and troughs come from. Photon pressure supplies force: $\nabla p_{\gamma}=\nabla \rho_{\gamma}/3,\quad \rho_{\gamma}\sim T^4$,
and the spatial variations of density themselves are also a source of changes in the gravitational potential.
 Potential and pressure gradients compete between them and induce acoustic oscillations in the fluid which result in 
temperature oscillations
$\Delta T\sim \delta \rho_{\gamma}^{1/4}\sim A(k) cos(kc_st)$, $(harmonic \; wave)$,
with $c_s$ the speed of sound. This photons do not find opposition either between their departure and their arrival
to their destination (our observing devices).
Photons released at maximum compression give blue spots as they have to climb up the potential, so they appear blue to us.
In contrast, those caught at maximum rarefaction form  red spots. In summary, the pattern of spots associated with 
long wavelength perturbations gets enriched by those coming from perturbations with smaller wavelengths. 

That of course, is not the end of the explanation, the angular power spectrum is the function describing how the variance of the fluctuations depends on the angular scale (its definition rests on the assumption of Gaussianity and randomness of the fluctuations).  Fluctuation extrema correspond to peaks in power (even ones correspond to compression and odd ones correspond to rarefaction).  Randomness of the fluctuations implies we have them in all wavelenghts, and clearly and when we compare a fluctuation of a given wavelength with another one with half of that wavelength it is obvious that the time the second takes to complete an oscillation is half of the time the first one takes. 
Thus, if by recombination there is a mode which has completed half an oscillation, there will modes which will have completed $2,3,4,\dots$ half oscillations. 

You may have noticed, as well, that the spectrum presents a modulation, it is due to different effects (see  \cite{Hu:2001bc}
 for an authorized discussion). Baryons being massive have preference for the throughs and the compression gets enhanced, thus the amplitude of odds peaks gets larger.
Another effect of baryons is making 
the oscillations slower so the peaks are pushed to higher multipoles.
On the other hand, photons are also important in the modulation as  in the radiation dominated phase. 
During radiation domination most gravity comes from  photons, they are the major source of gravitational potential 
so as pressure redistributes the photons,  gravitational potentials get washed away. This effect is absent when the Universe becomes matter dominated, so it is exclusive to high frequency modes, whereas low frequency modes do not start to oscillate till matter domination has begun.
Finally, there is also a strong damping effect due to slight imperfections in the fluid associated with shear viscosity and heat conduction (Silk damping \cite{silk}).

Interestingly, the CMB provides a dark energy test which requires considering only the background geometry, in contrast to other uses which rest on  the powerful but challenging study of perturbations.
This is the CMB shift test, which estimates how a physical length in the primordial universe appears to us.
By comparing the (non-Euclidianity) effects on geometry of different matter/contents  inference about the likeliest ones can be done.
The basic quantity to consider is that in an arbitrary  model the position of the first peak in the temperature spectrum is $l_1=\pi{D_A(z_{\rm rec})}/{r_s(z_{\rm rec})}$, 
where the quantity $r_s(z_{\rm rec})$ represents the last scattering sound horizon scale i.e. the distance sound may 
travel before the recombination epoch.

The CMB shift gives the first CMB peak position ratio between a model one wants to test (unprimed model) and a reference Einstein-de Sitter model (primed model)
$$R\equiv2{l_1}/{l_1^{'}}$$ \cite{bet}
It is considered as 
a robust test as it does not depend on  
the parameter $H_0$ (the Hubble factor today), which is not constrained by the other tests.
Considering the speed of sound $c_s$ is constant and using the approximations
\begin{equation}D'_A(z_{\rm rec})\approx{2c }{a_{\rm rec}}/{H_0} \; r_s(z_{\rm rec})\approx
2{c_s }{a^{3/2}_{\rm rec}}/{H_0\sqrt{\Omega_m}},\end{equation}
one finally arrives at
\begin{equation}
R\approx H_0\sqrt{\Omega_m}\int_0^{z_{\rm rec}}{dz}/{H(z)}.
\end{equation}
You can go by the value $R(1089)=1.71\pm0.03$ calculated from WMAP3 data. 

In a fashion similar to the other tests one will have to construct a $\chi^2$ function using the latter observational value
and the theoretical function for the cosmological model to be constrained.

In the next section we are going to consider a test which also stems from early universe physics.
\section{Baryon acoustic oscillations}
Diagnosing the mystifying new physics causing acceleration requires more
precise measurements of cosmological distance scales.
In recent years a new geometrical constraint of dark energy has emerged which relies on traces left by early universe sound waves in the galaxy distribution. This test is somehow in its infancy but it is very promising.
As it often in science, quite a few years have elapsed form prediction of the effect \cite{Bashinsky:2000uh,Sunyaev:1970eu}
till its detection \cite{Eisenstein:2005su}.

Let me move on now to a physical description of the phenomenon based on the one found in 
\cite{baopage}
(see as well \cite{eisenrev,White:2005tf}).
The early universe was composed of a plasma of energetic 
photons and ionized hydrogen (protons and electron) in addition to other trace elements.
Imagine now a single perturbation in the form of a excess of matter.
Pressure is very high and ejects the baryon-photon outward at relativistic velocities.
At first photon and baryons go by the hand (the speed of the radius of the shell being  larger than half  the speed of light).
 At recombination photons decouple and stream away whereas the baryon peak gets frozen; the reason for this is they are cold dark matter and therefore they have no intrinsic motion (no pressure) unlike the previous stages where they were coupled to the photons (in this early situation baryons were subject to radiation pressure).
The photon distributions become more and more homogeneous, while the baryon overdensity does not disappear.
Finally, the initial large gravitational potential well begins to attract matter back into it and a second overdense region appears at the center. Two stages in the evolution of a single perturbation are illustrated in Fig. 6

\begin{figure}[bt!]
\begin{tabular}{c}
\includegraphics[width=0.7\textwidth]{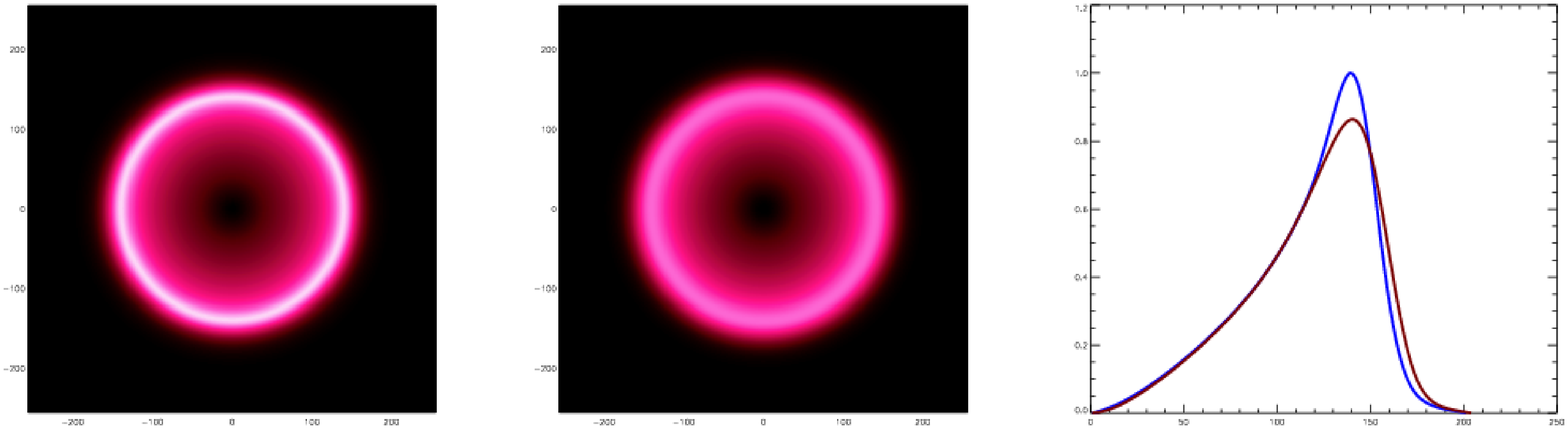}\\
\includegraphics[width=0.7\textwidth]{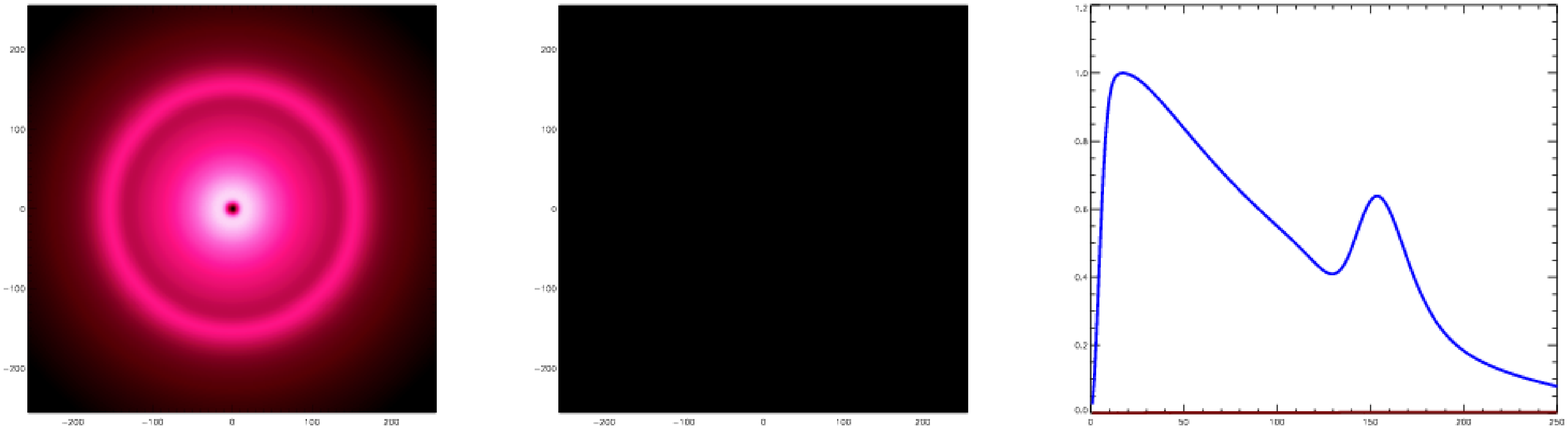}
\end{tabular}
\caption{Two stages in the evolution of a primordial density perturbation. Baryon density is on the left panels, photon density middle ones and mass profile in the right panel. The upper image is for a situation not much later than the decoupling time, the last one correspond to much later epoch. Images courtesy of Martin White.}
\end{figure}

The most important fact of the phenomenon is it leaves traceable effects on large-scale structure.
Initial perturbations would have produced wavy excitations emanating from all points, and the subsequent evolution
would resemble what happens when dropping rocks in a pond.
The plasma shells would have today 150 Mpc (or 500 million light years) 
and galaxy formation in the locus of those shells  would be likelier.
In fact, there is a correlation between galaxies in the shells and their centers which result in the detectable effect
that galaxies are more likely to be separated by that distance than by larger or smaller ones.

In consequence, the large scale two-point correlation function presents a clear peak at 100$h^{-1}$ Mpc. The two point correlation $\xi$ function (see Fig. 7) controls the joint probability of finding two galaxies
centered within the volume elements $dV_1$ and $dV_2$ at separation r (the galaxy number density is n and $r_0$ is the characteristic clustering length):
$dP=n^2[1+\xi(r/r_0)]dV_1dV_2$
\begin{figure}
\includegraphics[width=0.45\textwidth]{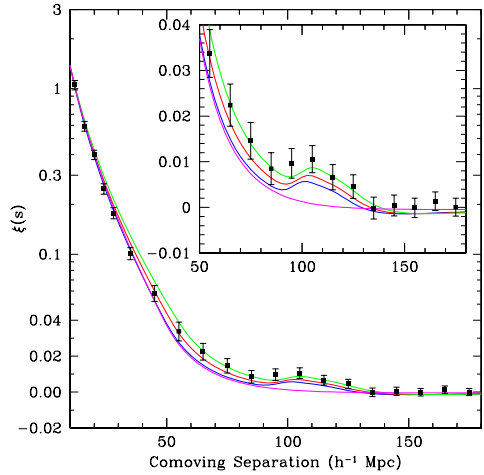}\\
\caption{Correlation function versus redshift. Image borrowed from \cite{Eisenstein:2005su}.}
\end{figure}
The derivation of the scale of the shell $r_{peak}$ involves postulating the form of $H(z)$, and as the value
of $r_{peak}$ predicted by CMB physics is known one can vary the parameters of the model to find the best fit.
The conversion between distances and redshifts can be done using the dilation scale $D_v(z)$. 
This quantity is an ``isotropic'' distance definition encoding the effects of expansion along the line of sight and along the transversal direction to it (see Fig. 2(b)):
\begin{eqnarray}
D_V(z)=[((P_3P_4)/\Delta\theta)^2(P_1P_2)]^{1/3}
\equiv (D_M(z)^2 \Delta D_C)^{1/3}
\end{eqnarray}
The newest results are those given in \cite{Percival:2007yw}.
Three galaxy catalogs were used: a main galaxy catalog combining 2dFGRS and SDSS main galaxies (this catalog fixes
$D_V(0.2)$), a catalog of SDSS large red galaxies (this catalog fixes $D_V(0.35)$), and a combination of the two. This combined sample gives $r_s/D_V(0.35)=0.1094\pm 0.0033$
and $r_s/D_V(0.35)=0.1980\pm 0.0058$. This data must be inserted in the adequate $\chi^2$ function and it will only remain to choose what model to test.

\section{Cosmologies with late-time acceleration}
There are, as you can imagine, a lot of cosmological models in the market with more or less physical foundation you could
try and apply the dark energy tests presented here. Basically, the lesson to be extracted from the previous discussion is that in principle just needs to postulate the functional form of $H(z)$. In what follows I wish to make an a outline of a few ot these models. This is, of course, a biased list of parameterizations as it is based on my own preferences, and on the other hand
there are authorized reviews you can head to for a more exhaustive revision.

In general, a given general relativistic  model of dark energy  with equation of state $w_{de}(z)$ leads to
\begin{equation}{H^2(z)}/{H_0^2}=\Omega_m(1+z)^3+(1-\Omega_m)\exp\left[3\int_0^z({1+w_{de}(x)})/({1+x})dx\right].
\end{equation} To get the latter a universe containing  dark energy and  dust (dark matter and baryons) has been assumed.

The first model I want to consider here is the {\bf LCDM model}.
It follows from the choice $w_{de}(z)=-1$, i.e.  
dark energy is a cosmological constant. 
Thus
\begin{equation}
{H^2(z)}/{H_0^2}=\Omega_m(1+z)^3+1-\Omega_m,
\end{equation}
with $\Omega_m$ a free parameter.
LCDM is consistent with all data, but there is a theoretical problem in explaining the observed value of the cosmological constant $\Lambda$.

The second case of this list is the {\bf DGP model}
It was proposed by Deffayet \cite{deffayet} the inspiration being \cite{dgp} 
 It represents a simple alternative to the standard LCDM Cosmology, with the same number of parameters. Late-time acceleration in this model is
due to an infrared modification of gravity (no dark energy needed). 
Explicitly one has
\begin{equation}
{H(z)}/{H_0}=({1-\Omega_m})/{2}+\sqrt{(1-\Omega_m)^2/{4}+\Omega_m(1+z)^3}.
\end{equation}

Let me now tell you about the {\bf QCDM model}.
It is the simplest generalization of 
the LCDM model. It consists in taking a constant value of $
w_{de}=w$ different in general from $-1$, so 
\begin{equation}
{H^2(z)}/{H_0^2}=\Omega_m(1+z)^3+(1-\Omega_m)(1+z)^{3(1+w)}.
\end{equation}
Even though it is a rather simple model, it can be useful for detecting at a first instance preference for
 evolutionary dark energy. 

Another two parameter model of interest is the {\bf LDGP model} \cite{ldgp}.
The  DGP model has actually two separate branches 
They reflect the two ways to embed the 4D brane universe in the 5D bulk spacetime.
We have loosely called  before DGP before 
to the self-accelerating branch (this is common practice in the literature) but I insist there is another branch which
is physically interesting too. The LDGP model precisely represents this branch, i.e. the non self-accelerating one, with a cosmological constant included which is required for acceleration.
Interestingly, higher values of the cosmological constant are allowed as compared to LCDM (due to an screening effect).
In a explicit way
\begin{equation}{H(z)}/{H_0}=\sqrt{\Omega_m(1+z)^3+1-\Omega_m+2\sqrt{\Omega_{r_c}}+\Omega_{r_c}
}-\sqrt{\Omega_{r_c}}
\end{equation}
and the parameter $\Omega_{r_c}$ encodes the so called crossover scale signaling the transition from the general relativistic to the modified gravity regime

The last but one case I wish to address is {\bf Chevallier-Polarski-Linder Ansatz} \cite{cpl}.
This is a widespread  generalization of
QCDM for which $$w_{de}(z)=w_0 + w_1 (1-1/(1+z)).$$
Therefore
\begin{equation}
{H^2(z)}/{H_0^2}=\Omega_m(1+z)^3
+(1-\Omega_m)(1+z)^{3(1+w_0+w_1)}e^{-3 {w_1 z}/{1+z}}.
\end{equation}
Two nice properties of it stand out, firstly the model displays finiteness of $H(z)$ at high redshifts, secondly it admits a simple physical interpretation as $w_1$ is a measure of the scalar field potential slow roll factor $V'/V$  in a quintessence picture \cite{linder}

Finally, to put and end on this list, I would like to consider the
{\bf QDGP model} \cite{qdgp}.
 This is a 
generalization of LDGP: the cosmological constant is replaced by dark energy with a constant equation of state $w$. The modified 4D Friedman equation has the following form:
\begin{eqnarray}{H(z)}/{H_0}&=&\sqrt{\Omega_m(1+z)^3+(1-\Omega_m+2\sqrt{\Omega_{r_c}})(1+z)^{3(1+w)}+\Omega_{r_c}
}-\sqrt{\Omega_{r_c}}.\end{eqnarray}

A remark about three of the models considered is in order. Models with late-time acceleration due to infrared modifications of gravity are called dark gravity models, and DGP, LDGP and QDGP are of that sort.
An effective dark energy equation of state $w_{\rm eff}$ can be deduced by imposing 
$\dot{\rho}_{\rm eff}+3H(1+w_{\rm eff})\rho_{\rm eff}=0$.
Calculating this in terms of redshift for the three models considered here can be a nice exercise for the readers.

I finish here this section as a turn of subject is required, as I feel this contribution would not be sort of self-contained if the issue of the statistical treatment of the tests considered was not covered.

\section{Statistical infererence}
Science, and ergo Astronomy, is about making decisions.
All scientific activities (design of experiments, design and building of instruments, data collection, reduction and interpretation, ...) rely on decisions. In turn, decisions are made by comparison, and comparison requires a statistics (a summarized description) of the 
available data.
The term ``statistics'' means rigorously a quantity giving  broad representation of the data, but the same term is also loosely used when referring to ``statistical inference''.
Moreover, as in science making decisions involves  measurements and derivation of values of parameters, it is necessary to assess the degree of belief of the true value of the parameter being measured/derived.    

Astronomy/Cosmology being about the study of the Universe makes them singular in what concerns decision making.
The Universe is not an experiment one can rerun, and to make things worse the typical objects studied are
very distant.
In consequence, typically one will have very poor knowledge of the distribution behind the variables subject to measurement.
Yet the difficulties must be overcome as statistics enters all five stages in the loop characterizing every experiment: 
observe-reduce-analyse-infer-cogitate.
 
I am going  to introduce now some basic notions, definitions and notation to be used here after. "The readers interested in learning more about most of the 
topics addressed in this section
can head to 
\cite{wall}, but I also recommend to you other sources such as \cite{justhe,robthe,liddle}.

Probability will be a numerical account of our strength of belief.
Let us enunciate now Kolmogorov axioms:
for any random event $A$ one has $0<prob(A)<1$, if and event $A$ is certain then $prob(A)=1$, if event $A$ and $B$ are exclusive, then $prob(A\, or \, B)=prob(A)+prob(B)$.
These axioms are basically all is needed to develop entirely the mathematical probability theory (by this one means the recipes to manipulate probabilities once they have been specified).
Probabilities are not inherent properties of physical problems, that means they do not make sense on their own, they are just a reflection of the knowledge we have.

If the knowledge about even $A$ does not affect the probability of event $B$, those events are independent, i.e. 
$prob(A\,and\,B)=prob(A)prob(B)$.
If $A$ and $B$ are not independent events then it is convenient to know conditional probabilities:
$prob(A\vert B)={prob(A\,and\,B)}/{prob(B)}$ and $prob(B\vert A)={prob(A\,and\,B)}/{prob(A)}$, which can be derived from Kolmogorov axioms.
If $A$ and $B$ are independent, then
$prob(A\vert B)=prob(A)\quad prob(B\vert A)=prob(B)$
In addition, if event $B$ comes is different flavours $(B_1,B_2,B_3,...)$, then
$p(A)=\sum_i prob(A\vert B_i)prob(B_i)$. Moreover, if $A$ is a parameter of insterest but the $B_i$ are not, knowledge of $prob(B_i)$ allows getting rid of those nuisance parameters by summation/integration, and the process is called marginalization.  

Having presented this basics let me guide you into the realm of Bayesian inference
By equating $prob(A\, and\, B)$ and $prob(B\, and\, A)$ one gets the identity
$prob(A\vert B)={prob(B\vert A)prob(B)}/{prob(A)} $
known as Bayes theorem.

There is a state of belief, the prior $prob(B)$, before the data, the event $A$, are collected, and experience modifies this knowledge.
Experience is encoded in the likelihood, $prob(A\vert B)$. The state of belief at the end of the process (analysis of the data)  is represented by the posterior $prob(B\vert A)$.
Mathematically na\"ive though it is, this theorem is, a regards interpretation, very powerful, but not devoid of controversy. 
I provide here only a short account of Bayesian statistics,  if you feel like
learning more about it, head to this site \cite{loredo}.

For a better understanding of the different between the Bayesian and frequentist approaches, consider now taking out balls from a box (you cannot see the inside and only know they are either  white or red).
Imagine you extract several times three balls and put them back into the box.
Would you care of the probability of taking out two whites and one red?
Would you rather not be more interested in saying something about the contents of the box?
The second question seems  the more natural one. This is the kind of question Bayes theorem allows to answer.
Bayesians care for probabilities of hypotheses given data, whereas frequentists are concerned about the 
probability of hypothetical data assuming the truth of some hypothesis. 

A useful concept in connection with all this is that of probability density functions.
If $x$ is a random real variable expressing a result, the probability of getting a number near x is 
$p(x)=prob(x)\delta x$,  
and $p(x)$ is the probability density function distribution of $x$ (probability density functions are also called probability distributions).   Probability density functions satisfy these properties:
$prob(a<x<b)=\int_a^bp(x)dx$,
$\int_{-\infty}^{\infty}p(x)dx=1$,
$p(x)$ is single-valued and non-negative for all real $x$.
Probability density functions are usually quantified by
the position of the center or mean $\mu=\int_{-\infty}^{\infty}xp(x)dx$,
the spread $\sigma^2=\int_{-\infty}^{\infty}(x-\mu)^2p(x)dx$.

Among them one stands out notoriously; it is the Gaussian probability density function, which is an ubiquitous distribution in Physics, thanks to the central limit theorem, which makes it work is most
situations: broadly speaking a little averaging makes any distribution converge to the Gaussian one,
explicitly, $p(x)=e^{-{(x-\mu)^2}/{\sigma}}/{\sqrt{2\pi\sigma}}$.

We are going to make use of it for parameter estimation and model selection, which is at the core of many investigations in Cosmology these days.
Consider we have $N$ observational data for some physical quantity $f^{th}$ of interest. We set
$\{f_j^{obs}\}=\{d_j\}$.
Assume $f^{th}$  depends on parameters $\{\theta_i\}$ and  consider it  in the context of a given model ${\cal M}$.
In addition, regard the errors as Gaussian distributed so the likelihood reads
\begin{equation}
{\cal L}(\{d_j\}\vert\{\theta_i\},{\cal M})\propto e^{-\chi^2({\{\theta_i\}})}\; with \; \chi^2({\{\theta_i\}})=\sum_{j=1}^N{[f_j^{obs}-f^{th}({\{\theta_i\}})]^2}/{\sigma_j^2}.\end{equation}
For uncorrelated observational datasets
\begin{eqnarray}
{\cal L}(\{d^{(1)}_j\}\cap\dots\cap\{d^{(m)}_k\}\vert \{\theta_i\}\,{\cal M})\propto{\cal L}(\{d^{(1)}_j\}\vert \{\theta_i\}\,{\cal M})\times\dots\times
{\cal L}(\{d^{(m)}_k\}\vert \{\theta_i\}\,{\cal M}).\end{eqnarray}

The probability density function $p(\{\theta_i\}\vert\{d_j\}, {\cal M})$  of the parameters to have values
$\{\theta_i\}$  under the assumption that the true  model is ${\cal M}$  and provided that the available observational data are $\{d_j\}$ is given by Bayes theorem.
In terms of probability density functions
\begin{equation}p(\{\theta_i\}\vert\{d_j\},{\cal M})= \frac{{\cal L}(\{d_j\}\vert\{\theta_i\},{\cal M})\pi(\{\theta_i\},{\cal M})}{\int{\cal L}(\{d_j\}\vert\{\theta_i\},{\cal M})\pi(\{\theta_i\},{\cal M})d\theta_1\dots d\theta_n}\,.
\end{equation}
Here
$p(\{\theta_i\}\vert\{d_j\},{\cal M})$  is called 
the posterior probability density function. 
and best fit values of the parameters 
are estimated by maximizing it, whereas 
$\pi(\{\theta_i\},{\cal M})$ is called 
the prior probability density function.
Choice of prior is subjective but compulsory in the Bayesian approach; remember the prior pdf encodes all previous knowledge about the parameters before the observational data have been collected.

The first step to estimate parameters in the Bayesian framework is
maximizing the posterior $p(\{\theta_i\}\vert\{d_j\},\vert{\cal M})$.
The second step is construction credible intervals 
It is convenient to simplify our notation, so 
$p(\{\theta_i\}\vert\{d_j\},{\cal M})\equiv p(\theta_1,\dots,\theta_n)$.
The marginal probability density function on $\theta_i$ is
${p}(\theta_i)=\int {p}(\theta_1,\dots,\theta_n)d\theta_1\dots d\theta_{i-1}d\theta_{i+1}\dots d\theta_{n}$.
Of course, if the model is genuinely uniparametric no marginalization is required.
It is also convenient to define parameters $\theta_{il}$ and $\theta_{iu}$ satisfying
$p(\theta_{il})\simeq0$ and $p(\theta_{iu})\simeq0$
The credible intervals, under the hypothesis that the marginal pdf is approximately Gaussian, are constructed 
from the median and errors. The $68\%$ percent credible intervals on the parameter $\theta_i$ are given as $\theta_i=x_{-y}^{+z}$

The median $x$ is calculated from 
$\int_ {\theta_{il}}^{x}{p}(\theta_i)d\theta_i=0.5\times\int{p}(\theta_i)d\theta_i$
The lower error  $y$ is calculated from 
$\int_{\theta_{il}}^{x-y} p(\theta_i)d\theta_i=((1-0.68)/2)\times\int{ p}(\theta_i)d\theta_i\quad$
The upper error  $z$ is calculated from 
$\int_{x+z}^{\theta_{iu}}p(\theta_i)d\theta_i=((1-0.68)/2)\times\int{p}(\theta_i)d\theta_i$
 
Credible contours are a popular/illustrating construction and they are worth a mention.
By marginalization with respect to all parameters but two one gets
$$p(\theta_i,\theta_{i+1})=
{p}(\theta_i)=\int {p}(\theta_1,\dots,\theta_n)d\theta_1\dots d\theta_{i-1}d\theta_{i+2}\dots d\theta_{n}$$ 
Again, obviously, marginalization is not necessary  if the model is genuinely biparametric.
An effective $\chi^2$ can be defined in the form $-2\log(p(\theta_i,\theta_{i+1}))=\chi^2_{eff}(\theta_i,\theta_{i+1})/2$ (up to a constant),
so the best fit responds to the maximum posterior criterion.
The probability that for some parameters other than those for the best fit (maximum likelihood)
 $\chi^2$ increases  
 with respect to the best fit by an amount $\Delta \chi^2$ is  $1-\Gamma(1,\log (p_{bf}/p))/\Gamma(1)$. By fixing the desired probability content, 
 the latter becomes
 the implicit equation of the credible contours
 on the parameter space  (popular choices in the literature are $68.3\%$, $95.4\%$ and $99.7\%$).

Now let me comment about model selection. Bayesians use a estimator to select models
which  informs about how well the parameters of the model
fit the data. It does not rely exclusively on the best-fitting parameters of the model,
but rather it involves an averaging over all the parameter values that were theoretically plausible before the measurement ever took place \cite{liddle}.
Bayes evidence is
\begin{equation}
{\cal E({\cal M})}= p(\{d_j\}\vert {\cal M})=\int\pi(\{\theta_i\},{\cal M}){\cal L}(\{d_j\}\vert\{\theta_i\},{\cal M}) d\theta_1\dots d\theta_n,
\end{equation}
so it is the probability of data  $\{d_j\}$ given the model ${\cal M}$
Here $\pi(\{\theta_i\},{\cal M})$ is the model's prior on the set of parameters, normalized to unity (i.e. $\int \pi(\{\theta_i\},{\cal M})d\theta_1\dots d\theta_n\!\!=1\!$.)
Using the popular top-hat prior $\pi(\theta_i)=({\theta_{i {\rm max}}-\theta_{i {\rm min}}})^{-1}$ Bayes evidence is rewritten as
$${\cal E(M)}={\left(\int_{\theta_{1 {\rm min}}}^{\theta_{1 {\rm max}}}\dots \int_{\theta_{n {\rm min}}}^{\theta_{n {\rm max}}} {\cal L}(\theta_1,\dots,\theta_n)d\theta_1\dots \theta_n\right)}/{\left(\int_{\theta_{1 {\rm min}}}^{\theta_{1 {\rm max}}}\dots \int_{\theta_{n {\rm min}}}^{\theta_{n {\rm max}}} d\theta_1\dots d\theta_n\right)}.\nonumber
$$
Finally, preference of model ${\cal M}_i$ over model ${\cal M}_j$ given $\{d_k\}$ is estimated by
$$
{p({\cal M}_i\vert\{d_k\})}/{p({\cal M}_j\vert\{d_k\})}={{\cal E}_i ({\cal M}_i)}{\pi_i({\cal M}_i)}/{{\cal E}_j({\cal M}_j)}{\pi_j({\cal M}_j)}$$
The Bayes factor $B_{ij}$ for any two models ${\cal M}_i$ and ${\cal M}_j$ is 
$$
B_{ij}={{\cal E}_i ({\cal M}_i)}/{{\cal E}_j ({\cal M}_j)}
$$
so if as in usual practice one assumes 
no prior preference of one model over the other, that is, 
Assuming $\pi_i({\cal M}_i)=\pi_j({\cal M}_j)=1/2$ (no {\it a priori} preference) then 
$$
{p({\cal M}_i\vert\{d_k\})}/{p({\cal M}_j\vert\{d_k\})}=B_{ij}.
$$
The most popular key to interpreting Bayes factors is Jeffreys scale \cite{jeffreys}:
if $ln(B_{ij})<1$, then the evidence against ${\cal M}_j$ is not significant, if $1<ln(B_{ij})<2.5$, then the evidence against ${\cal M}_j$ is substantial, if $2.5<ln(B_{ij})<5$, then the evidence in favor of ${\cal M}_i$ is strong, if $5<ln(B_{ij})$, then the evidence in favor of ${\cal M}_i$ is decisive.

There are of course many more things which could be mentioned about this topic, but I hope this condensed introduction will help you start crossing swords with the art of constraining dark energy models using geometrical tests like
the ones discussed here or upcoming ones.
 
\section{Conclusions}
In this contribution I have tried to review some of the background on geometrical tests of dark energy models. A historical review of the development of Cosmology has been followed by an account of the importance of the
discovery of the late-time acceleration, which is commonly attributed to the existence of an exotic component in the cosmic budget. Other possible explanations have been attempted, which required being open minded enough to admit the existence of extra dimensions.

Examples of both conventional (if that adjective can be used) and extradimensional models have been mentioned here with respect to their geometrical features. All these models can be subject to different observational tests which only require postulating a parametrization of the Hubble factor of the Universe or quantities derived from it. Here I have been concerned with four tests only: luminosity of supernovae, direct $H(z)$ measurements, the CMB shift and baryon acoustic oscillations. I cannot deny this selection is biased in the sense these are tests I have made research on, but the first and the last one are definitely very important and much effort is doing by research groups all over the world
in finding the data their application requires. 

This contribution has also tried to maintain a pedagogical tone as it has been prepared for the Advanced Summer School 2007 organized by Cinvestav in Mexico DF. I just hope it will be useful to any reader which happens to come across it. 

\section{Acknowledgments}
I wish to thank to Nora Bret\'on, Mauricio Carbajal and Oscar Rosas-Ortiz for giving me the opportunity
to present this contribution to the Advanced School Summer School in Physics 2007 at Cinvestav.
I am also much indebted to Elisabetta Majerotto, as I have borrowed part of the material presented in this lectures
I have borrowed from work done in collaboration with her. Finally, I wish to thank 
Mariam Bouhmadi and again  Nora Bret\'on for reading the manuscript and helping improving it and to Ra\'ul B. P\'erez-S\'aez for a lovely picture.

\end{document}